\documentclass{aa}    

\usepackage{times}                          
\usepackage{txfonts}
\usepackage{booktabs}
\usepackage{longtable}
\usepackage{caption}
\usepackage{subcaption}

\newcommand{\nspecies}{14}  
\newcommand{\nrxns}{76}  

\newcommand{\cobold}{\texttt{CO$^5$BOLD}}

\newcommand{\rAA}{\ensuremath{\mathrm{\AA}}}

\newcommand{\mol}[2]{\ensuremath{\mathrm{#1}_#2}}  
\newcommand{\CT}{\mol{C}{2}}

\newcommand{\subrm}[1]{_\mathrm{#1}}

\newcommand{\comment}[1]{}

\newcommand{\matr}[1]{\mathbf{#1}}

\newcommand{\Teff}{T_{\mathrm{eff}}}

\newcommand{\qunit}[2]{{#1}\,\mathrm{#2}}  
\newcommand{\kms}{km s$^{-1}$}

\newcommand{\wlAA}[1]{\qunit{#1}{\rAA}}

\usepackage{graphics,graphicx,epsfig}   
\graphicspath{{figs/}}

\usepackage{amsmath}                                            

\usepackage{etoolbox}                       
\makeatletter
\patchcmd{\frontmatter@RRAP@format}{(}{}{}{}
\patchcmd{\frontmatter@RRAP@format}{)}{}{}{}
\makeatother

\usepackage{fancyhdr}
\usepackage{gensymb}

\usepackage{wrapfig}
\usepackage{import}

\usepackage[compact]{titlesec}         
\titlespacing{\section}{0pt}{2pt}{2pt} 

\begin{document}

\title{Implications of time-dependent molecular chemistry in metal-poor dwarf stars}
\date{Submitted: \today{}}
\author{S.A.\,Deshmukh \inst{1}
  \and H.-G.\,Ludwig\inst{1}
}

\institute{
  Zentrum f\"ur Astronomie der Universit\"at Heidelberg,
  Landessternwarte, K\"onigstuhl 12,
  69117 Heidelberg, Germany\\
  \email{sdeshmukh@lsw.uni-heidelberg.de}
  \fnmsep
}

\date{Received 24/01/2023; accepted 18/04/2023}

\abstract
{Binary molecules such as CO, OH, CH, CN, and \CT\ are often used as abundance indicators
  in stars. These species are usually assumed to be formed in chemical equilibrium.
  The time-dependent effects of hydrodynamics can affect the formation and
  dissociation of these species and may lead to deviations from chemical
  equilibrium.}
{We aim to model departures from chemical equilibrium in dwarf stellar
  atmospheres by considering time-dependent chemical kinetics alongside
  hydrodynamics and radiation transfer. We examine the effects of a decreasing
  metallicity and an altered C/O ratio on the chemistry when compared to
  the equilibrium state.}
{We used the radiation-(magneto)hydrodynamics code \cobold\, and its own chemical
  solver to solve for the chemistry of $\nspecies$ species and $\nrxns$ reactions. The
  species were treated as passive tracers and were advected by the velocity
  field. The steady-state chemistry was also computed to isolate the effects of
  hydrodynamics.}
{In most of the photospheres in the models we present, the mean
  deviations are smaller than $0.2$ dex, and they generally appear above
  $\log{\tau} = -2$. The deviations increase with height because the
  chemical timescales become longer with decreasing density and temperature.
  A reduced metallicity similarly results in longer chemical timescales and in a
  reduction in yield that is proportional to the drop in metallicity; a decrease by a factor $100$
  in metallicity loosely corresponds to an increase by factor $100$  in  chemical timescales. As both CH and OH are formed along reaction pathways to
  CO, the C/O ratio means that the more abundant element gives faster
  timescales to the constituent molecular species. Overall, the carbon enhancement
  phenomenon seen in very metal-poor stars is not a result of an improper
  treatment of molecular chemistry for stars up to a metallicity as low as
    [Fe/H] = $-3.0$.}
{}

\keywords{
  Stars: abundances --
  Stars: chemically peculiar --
  Stars: atmospheres --
  Hydrodynamics
}

\maketitle
%

\section{Introduction}
\label{sec:intro}

Stellar atmospheres are generally assumed to preserve the makeup of their birth
environment. The abundance of elements heavier than helium (known as metals) in
a stellar atmosphere is an indication of the stellar age, with older stars being
deficient in metals.
Spectroscopy is one of the foremost tools in determining the abundances of
various elements in stellar atmospheres. Since the first studies on solar
abundances in
the 1920s \citep{payneStellarAtmospheres1925, unsoeldUeberStruktur1928,
  russellCompositionSun1929} to modern large-scale
surveys such as the Gaia-ESO survey \citep{gilmoreGaiaESOPublic2012}, \textit{Gaia} \citep{gaiacollaborationGaiaData2022}, GALAH
\citep{bland-hawthornGALAHSurvey2016}, and \textit{Pristine}
\citep{starkenburgPristineSurvey2017}, to name a few, spectroscopically
determined stellar parameters have been a key tool in understanding the
composition of stellar atmospheres. Instrumentation
and modelling have been refined in tandem, with improvements such as the
treatment of departure
from local thermodynamic equilibrium (LTE) and advancements in one-dimensional (1D)
and three-dimensional (3D) model atmospheres. These directly lead to improvements
in the determination of solar and stellar abundances because the methods for doing so often
rely on model atmospheres and the assumptions therein. As a core component of
Galactic archaeology, abundance determinations of stellar photospheres from spectroscopy
often assume chemical equilibrium (implicitly assumed
within the LTE assumption). While LTE studies have been used historically to
determine stellar abundances \citep{asplundLineFormation2000,
  holwegerPhotosphericAbundances2001a, caffauSolarChemical2011}, the
accurate treatment of the departure from LTE of level populations (known as
radiative NLTE treatment) has been shown to provide more accurate abundances
in both solar and stellar photospheres
\citep{bergemannRedSupergiant2013,
  wedemeyerStatisticalEquilibrium2001,amarsiCarbonOxygen2019,
  mashonkinaNonlocalThermodynamic2020, maggObservationalConstraints2022}.

Molecular features are important in metal-poor (MP) stars because atomic lines are
comparatively weak
\citep{beersSearchStars1992, aokiHighresolutionSpectroscopy2013,
  yongMOSTMETALPOOR2013, kochPurveyorsFine2019}. In recent
years, increasingly metal-poor stars have been
discovered \citep{beersSearchStars1992, beveridgeChemicalCompositions1994,
  aokiHighresolutionSpectroscopy2013, hughesGALAHSurvey2022} , with a tendency of
a carbon enhancement in their atmospheres
\citep{beersDiscoveryAnalysis2005, sivaraniFirstStars2006,
  carolloCARBONENHANCEDMETALPOOR2014, cohenFrequencyCarbon2005,
  hansenAbundancesCarbonenhanced2016, luceyMillionCarbonEnhanced2022}.
These carbon-enhanced metal-poor (CEMP) stars comprise a large fraction of the
low-metallicity tail of the metallicity distribution function in the Galactic
halo \citep{norrisHE055748402007, susmithaOxygenAbundances2020}. Although NLTE
treatment of spectral lines is becoming
more prominent \citep{bergemannRedSupergiant2013, bergemannObservationalConstraints2019, mashonkinaNonlocalThermodynamic2020}, most of the work
concerning these abundance determinations is still done under the assumption of
chemical equilibrium, that is, that all chemical species are in equilibrium with one
another. Most NLTE studies consider radiative NLTE, meaning that the radiation
field is not in equilibrium with the local background temperature. This changes
the population of energy levels in an atom or molecule.
Radiative NLTE is still considered in a time-independent fashion. We instead
model the time-dependent chemical processes for a variety of species to
investigate the effects of hydrodynamics on molecular formation and dissociation
to study whether the carbon enhancement seen at very low metallicities is a real
effect or is due to a lack of consideration for time-dependent chemistry.

Chemical species will react with one another in such a manner as to approach
thermodynamic equilibrium, given enough time. However, as the rates of these
reactions depend strongly on temperature and density \citep{hornGeneralMass1972},
there may be regions in
the star in which chemical equilibrium conditions are not met. In the deeper,
hotter, collision-dominated layers, chemical species evolve to equilibrium
on timescales much faster than other physical timescales in the system. The
assumption of chemical equilibrium therefore implies that the chemistry evolves
to its equilibrium state faster than other processes can significantly perturb
it. In this work, the other physical processes are hydrodynamical, and the key
question is whether the chemistry reaches its local thermodynamic equilibrium
before the species are advected. Convection in a stellar atmosphere can also
lead to compression shocks, which quickly heat material. When chemical
kinetics are coupled to these processes, the chemistry evolves on a finite timescale, and
a prevalence of these hydrodynamical effects can push the overall chemistry out
of its local equilibrium state.

Metallicity also has a large impact on both the overall structure of
the atmosphere and the number densities of the species. At a cursory glance,
reducing the metallicity by a factor of $100$ immediately results in
a reduction by a factor of $100$ in the number densities, which naturally
results in slower
mass-action reaction rates. Relative abundances (especially of C and O) also
play a large role in determining the final yield as well as the chemical
timescales of different species \citep{hubenyTheoryStellar2015}. As a result of
the mass-action rates alone then, the sharp reduction in chemical
timescales may cause the chemistry to become out of equilibrium in higher,
cooler layers.

Currently, many different codes exist for modelling stellar atmospheres.
While 1D atmospheres have been used to great effect
\citep{gustafssonGridMARCS2008, allardModelAtmospheres1995},
3D time-dependent modelling is essential for
accurately modelling hydrodynamical effects within an atmosphere
\citep{pereiraHowRealistic2013}. Codes such as
\cobold\, \citep{freytagSimulationsStellar2012}, Stagger
\citep{magicStaggergridGrid2013}, Bifrost \citep{gudiksenStellarAtmosphere2011},
MuRAM \citep{voglerSimulationsMagnetoconvection2005}, and Mancha
\citep{khomenkoNumericalSimulations2017}
are prominent examples. We used \cobold\ model atmospheres to model
hydrodynamics, radiation transfer, and time-dependent chemistry together.

We investigate two distinct methods to treat the chemical evolution in a stellar
atmosphere. The first is to evolve the chemistry as a post-processing step,
using outputs from model atmospheres (known as snapshots) in
order to determine the chemical evolution of various species.
This method yields accurate results in regimes where the density-temperature
profile is conducive to fast-evolving chemistry (in comparison to advection).
The second method is to evolve the chemistry alongside the hydrodynamics, which is usually done
after advecting the species. While this is much more expensive computationally,
it will yield accurate results even in regimes in which the timescales of the
chemistry are comparable to the advection timescale. In principle, both
approaches are equivalent given a fine enough cadence because the chemical
species are treated as passive scalars. In other words, given a fine enough
sampling of snapshots, the post-processing method would tend towards the full
time-dependent treatment. The post-processing method of evolving species into
equilibrium is hence an approximation; using the final abundances calculated
with this method as presented here implicitly assumes the formation of these
species in chemical equilibrium. It is precisely this assumption that we
investigate by comparing these two methods.

\citet{wedemeyer-bohmCarbonMonoxide2005} investigated CO in the solar
photosphere and chromosphere in 2D, employing a chemical network with $\text{7}$
species and $27$ reactions. \citet{wedemeyer-bohmFirstThreeDimensional2006}
then expanded this into a 3D
analysis, showing the formation of CO clouds at higher layers. We build
on this further to include an extended chemical network involving $14$ species
and $83$ reactions, and focus on the photospheres of main-sequence turn-off
dwarf stars.
We investigate CO, CH, \CT, CN, and OH in detail because these $5$
species are spectroscopically interesting for abundance determinations in MP
stars.

The numerical methods and chemical network setup are described in Sec.
\ref{sec:methods}. The results of the three-dimensional simulations for the
time-dependent and steady-state calculations are presented in Sec.
\ref{sec:results} and discussed in Sec. \ref{sec:discussion}. Our conclusions
are given in Sec. \ref{sec:conclusion}.
\section{Method}
\label{sec:methods}
\subsection{Chemical reaction network}
\label{subsec:crn}

The chemical reaction network (CRN) that describes the system of differential
equations builds on the network presented in
\citet{wedemeyer-bohmCarbonMonoxide2005}, extending it to
$\nspecies$ species and $\nrxns$ reactions. Table \ref{table:cno_rn}
describes these
reactions along with the parameters of the rate coefficients.
Our network is focused on the evolution of CO, CH, \CT, CN, and OH
through reactions with neutral atomic and bimolecular species. Radiative
association, species exchange, two- and three-body reactions,
and collisional dissociation are included. Each reaction is given in the
modified Arrhenius form, parametrised by the pre-exponential factor
$\alpha$, an explicit temperature dependence $\beta,$ and a characterisation
of the activation energy $\gamma$ (see Sec.~\ref{subsec:rate_coefficients}
for a more detailed explanation). Some reactions with CO are catalysed reactions
and include a characteristic metal M.

The choice of reactions is discussed below. Generally, the CRN
was built to analyse the species CO, CH, CN, \CT , and OH. As the network
presented in \citet{wedemeyer-bohmCarbonMonoxide2005} already includes a
good treatment of CO, CH, and OH, we supplement this network with reactions
taken from the UMIST Astrochemistry Database
\citep{mcelroyUMISTDatabase2013} to model the other molecular species. Only
neutral atomic and bimolecular species were considered due to their
prevalence compared to other trace molecules, and the storage limitations
imposed by considering a full 3D time-dependent treatment.
We neglect photodissociation in this network, but we accept that the
effects may not be negligible in very optically thin layers. Additionally,
as the reactions used here often come from studies in planetary
atmospheres and combustion chemistry,
the reactions we present are sometimes defined outside of their temperature
limits, especially when considering deep photospheric regions. We chose to
focus on higher cooler layers for this reason.

\textsf{Reaction 58.}
We chose to use the rate that includes only
$\alpha$, instead of the rate that includes explicit temperature
dependence. This is because the temperature limits of this reaction are
$10-300$ K, and including the temperature-dependent rate would lead to a
much greater extrapolation due to the comparatively high temperatures in
the model atmospheres.

\textsf{Reactions 116, 133, and 198.}
For each of these reactions, two rates are presented in the database for
temperature limits of $10-300$ K, and $300-3000$ K. We opted to use the
latter rate as the temperature limits are closer to our use case.

\textsf{Reaction 206.}
The reaction is defined for the temperature limits $298-3300$ K and
$295-4000$ K. We opted to use the latter rate, whichincludes a higher
upper-temperature limit.

\textsf{Reaction 236.}
The reaction is defined for the temperature limits $10-500$ K and
$158-5000$ K. We opted to use the latter rate, which includes a higher
upper-temperature limit.

\textsf{Reaction 244.}
The reaction is defined for the temperature limits $10-294$ K and
$295-4500$ K. We opted to use the latter rate, which includes a higher
upper-temperature limit.

\begin{figure}[h]
  \centering
  \scalebox{0.5}{
    \includegraphics[]{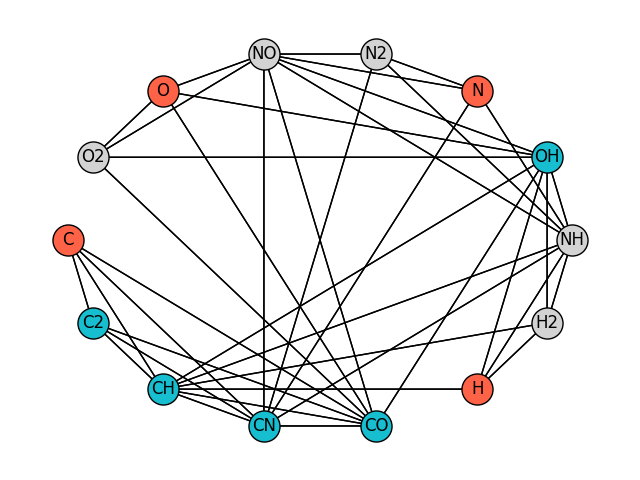}
  }
  \caption{
    Graph of the chemical reaction network with atoms (red), key molecular
    species (blue), and remaining molecular species (grey). The connections
    describe the reaction pathways qualitatively.
  }
  \label{fig:network_graph}
\end{figure}

A visualisation of the reaction
network is shown in Fig.~\ref{fig:network_graph}. Atomic species are shown in
red, key molecular species are shown in blue, and all other molecular species
are shown in grey. The full network with all reactions is too complex to show in
full detail, so that we chose to highlight the important reactions as edges between
nodes. The network is clearly connected, meaning that
any node can be reached starting from any other node, but it is not fully connected because every node
does not share an edge with every other node. These properties allowed us to find
reaction pathways in the reaction network (see
Sec.~\ref{subsec:timescales-pathways}).

\onecolumn
\begin{longtable}{r l c l r r r r}
  \caption{
    Reactions used in this work. ``Index'' refers to the index in the UMIST
    astrochemistry database. All reactions are of modified-Arrhenius form with
    a rate coefficient
    $k(T) = \alpha \left(\frac{T}{300}\right)^\beta \exp\left(\frac{-\gamma}{T}\right)$.
    The references are
    UMIST: \citet{mcelroyUMISTDatabase2013}.
    BDHL72: \citet{baulchEvaluatedKinetic1972}.
    KCD: \citet{konnov2000}.
    BDDG76: \citet{baulchEvaluatedKinetic1976}.    W80: \citet{westleyTableRecommended1980}.
  }                                                                                                                                          \\                                                                                                            \\                                                                                                                            \\
  \hline
  Index & Reactants                      &                          & Products                   & $\alpha$ & $\beta$ & $\gamma$ & Reference \\
  \hline
        &                                & Radiative Association    &                            &          &         &          &           \\
  \hline
  3681  & C   +             H            & $\Longrightarrow$        & CH       + $\gamma$        & 1.00e-17 & 0.00    & 0.0      & UMIST     \\
  3683  & H   +             O            & $\Longrightarrow$        & OH       + $\gamma$        & 9.90e-19 & -0.38   & 0.0      & UMIST     \\
  3703  & C   +             C            & $\Longrightarrow$        & C$_2$    + $\gamma$        & 4.36e-18 & 0.35    & 161.3    & UMIST     \\
  3705  & C   +             N            & $\Longrightarrow$        & CN       + $\gamma$        & 5.72e-19 & 0.37    & 51.0     & UMIST     \\
  3707  & C   +             O            & $\Longrightarrow$        & CO       + $\gamma$        & 1.58e-17 & 0.34    & 1297.0   & UMIST     \\
  3730  & O   +             O            & $\Longrightarrow$        & O$_2$    + $\gamma$        & 4.90e-20 & 1.58    & 0.0      & UMIST     \\
  \hline
        &                                & 3-body association       &                            &          &         &          &           \\
  \hline
  4079  & H   +             M   +  O     & $\Longrightarrow$        & M   +         OH           & 4.33e-32 & -1.00   & 0.0      & UMIST     \\
  4097  & C   +             M   +  O     & $\Longrightarrow$        & CO  +         M            & 2.14e-29 & -3.08   & -2114.0  & BDDG76    \\
  5000  & H   +             H   +  M     & $\Longrightarrow$        & H$_2$  +         M         & 6.43e-33 & -1.00   & 0.0      & KCD       \\
  5001  & H   +             H   +  H$_2$ & $\Longrightarrow$        & H$_2$  +         H$_2$     & 9.00e-33 & -0.60   & 0.0      & KCD       \\
  5002  & H   +             H   +  H     & $\Longrightarrow$        & H   +         H$_2$        & 4.43e-28 & -4.00   & 0.0      & BDHL72    \\
  7000  & H   +             H   +  O     & $\Longrightarrow$        & H   +         OH           & 1.00e-32 & 0.00    & 0.0      & BDHL72    \\
  7001  & C   +             H   +  O     & $\Longrightarrow$        & CO  +         H            & 2.14e-29 & -3.08   & -2114.0  & BDDG76    \\
  \hline
        &                                & Species Exchange         &                            &          &         &          &           \\
  \hline
  1     & CH  +             H            & $\Longrightarrow$        & C   +         H$_2$        & 2.70e-11 & 0.38    & 0.0      & UMIST     \\
  3     & H   +             NH           & $\Longrightarrow$        & H$_2$  +         N         & 1.73e-11 & 0.50    & 2400.0   & UMIST     \\
  8     & H   +             OH           & $\Longrightarrow$        & H$_2$  +         O         & 6.99e-14 & 2.80    & 1950.0   & UMIST     \\
  11    & C$_2$  +             H         & $\Longrightarrow$        & C   +         CH           & 4.67e-10 & 0.50    & 30450.0  & UMIST     \\
  14    & CO  +             H            & $\Longrightarrow$        & C   +         OH           & 5.75e-10 & 0.50    & 77755.0  & W80       \\
  18    & H   +             NO           & $\Longrightarrow$        & NH  +         O            & 9.29e-10 & -0.10   & 35220.0  & UMIST     \\
  19    & H   +             NO           & $\Longrightarrow$        & N   +         OH           & 3.60e-10 & 0.00    & 24910.0  & UMIST     \\
  24    & H   +             O$_2$        & $\Longrightarrow$        & O   +         OH           & 2.61e-10 & 0.00    & 8156.0   & UMIST     \\
  42    & C   +             H$_2$        & $\Longrightarrow$        & CH  +         H            & 6.64e-10 & 0.00    & 11700.0  & UMIST     \\
  44    & H$_2$  +             N         & $\Longrightarrow$        & H   +         NH           & 1.69e-09 & 0.00    & 18095.0  & UMIST     \\
  48    & H$_2$  +             O         & $\Longrightarrow$        & H   +         OH           & 3.14e-13 & 2.70    & 3150.0   & UMIST     \\
  52    & H$_2$  +             O$_2$     & $\Longrightarrow$        & OH  +         OH           & 3.16e-10 & 0.00    & 21890.0  & UMIST     \\
  58    & C   +             CH           & $\Longrightarrow$        & C$_2$  +         H         & 6.59e-11 & 0.00    & 0.0      & UMIST     \\
  61    & C   +             NH           & $\Longrightarrow$        & CH  +         N            & 1.73e-11 & 0.50    & 4000.0   & UMIST     \\
  62    & C   +             NH           & $\Longrightarrow$        & CN  +         H            & 1.20e-10 & 0.00    & 0.0      & UMIST     \\
  66    & C   +             OH           & $\Longrightarrow$        & CH  +         O            & 2.25e-11 & 0.50    & 14800.0  & UMIST     \\
  67    & C   +             OH           & $\Longrightarrow$        & CO  +         H            & 1.81e-11 & 0.50    & 0.0      & W80       \\
  68    & C   +             CN           & $\Longrightarrow$        & C$_2$  +         N         & 4.98e-10 & 0.00    & 18116.0  & UMIST     \\
  70    & C   +             CO           & $\Longrightarrow$        & C$_2$  +         O         & 2.94e-11 & 0.50    & 58025.0  & UMIST     \\
  71    & C   +             N$_2$        & $\Longrightarrow$        & CN  +         N            & 8.69e-11 & 0.00    & 22600.0  & UMIST     \\
  75    & C   +             NO           & $\Longrightarrow$        & CN  +         O            & 6.00e-11 & -0.16   & 0.0      & UMIST     \\
  76    & C   +             NO           & $\Longrightarrow$        & CO  +         N            & 9.00e-11 & -0.16   & 0.0      & UMIST     \\
  80    & C   +             O$_2$        & $\Longrightarrow$        & CO  +         O            & 5.56e-11 & 0.41    & -26.9    & UMIST     \\
  100   & CH  +             N            & $\Longrightarrow$        & C   +         NH           & 3.03e-11 & 0.65    & 1207.0   & UMIST     \\
  102   & CH  +             O            & $\Longrightarrow$        & C   +         OH           & 2.52e-11 & 0.00    & 2381.0   & UMIST     \\
  104   & CH  +             O            & $\Longrightarrow$        & CO  +         H            & 1.02e-10 & 0.00    & 914.0    & UMIST     \\
  116   & CH  +             O$_2$        & $\Longrightarrow$        & CO  +         OH           & 7.60e-12 & 0.00    & 0.0      & UMIST     \\
  126   & N   +             NH           & $\Longrightarrow$        & H   +         N$_2$        & 4.98e-11 & 0.00    & 0.0      & UMIST     \\
  130   & N   +             OH           & $\Longrightarrow$        & NH  +         O            & 1.88e-11 & 0.10    & 10700.0  & UMIST     \\
  131   & N   +             OH           & $\Longrightarrow$        & H   +         NO           & 6.05e-11 & -0.23   & 14.9     & UMIST     \\
  132   & C$_2$  +             N         & $\Longrightarrow$        & C   +         CN           & 5.00e-11 & 0.00    & 0.0      & UMIST     \\
  133   & CN  +             N            & $\Longrightarrow$        & C   +         N$_2$        & 1.00e-10 & 0.40    & 0.0      & UMIST     \\
  138   & N   +             NO           & $\Longrightarrow$        & N$_2$  +         O         & 3.38e-11 & -0.17   & -2.8     & UMIST     \\
  144   & N   +             O$_2$        & $\Longrightarrow$        & NO  +         O            & 2.26e-12 & 0.86    & 3134.0   & UMIST     \\
  195   & NH  +             NH           & $\Longrightarrow$        & H$_2$  +         N$_2$     & 1.70e-11 & 0.00    & 0.0      & UMIST     \\
  197   & NH  +             O            & $\Longrightarrow$        & N   +         OH           & 1.16e-11 & 0.00    & 0.0      & UMIST     \\
  198   & NH  +             O            & $\Longrightarrow$        & H   +         NO           & 1.80e-10 & 0.00    & 300.0    & UMIST     \\
  206   & NH  +             NO           & $\Longrightarrow$        & N$_2$  +         OH        & 1.46e-11 & -0.58   & 37.0     & UMIST     \\
  236   & O   +             OH           & $\Longrightarrow$        & H   +         O$_2$        & 1.77e-11 & 0.00    & -178.0   & UMIST     \\
  240   & C$_2$  +             O         & $\Longrightarrow$        & C   +         CO           & 2.00e-10 & -0.12   & 0.0      & UMIST     \\
  243   & CN  +             O            & $\Longrightarrow$        & C   +         NO           & 5.37e-11 & 0.00    & 13800.0  & UMIST     \\
  244   & CN  +             O            & $\Longrightarrow$        & CO  +         N            & 5.00e-11 & 0.00    & 200.0    & UMIST     \\
  251   & N$_2$  +             O         & $\Longrightarrow$        & N   +         NO           & 2.51e-10 & 0.00    & 38602.0  & UMIST     \\
  261   & NO  +             O            & $\Longrightarrow$        & N   +         O$_2$        & 1.18e-11 & 0.00    & 20413.0  & UMIST     \\
  377   & C$_2$  +             O$_2$     & $\Longrightarrow$        & CO  +         CO           & 1.50e-11 & 0.00    & 4300.0   & UMIST     \\
  382   & CN  +             CN           & $\Longrightarrow$        & C$_2$  +         N$_2$     & 2.66e-09 & 0.00    & 21638.0  & UMIST     \\
  387   & CN  +             NO           & $\Longrightarrow$        & CO  +         N$_2$        & 1.60e-13 & 0.00    & 0.0      & UMIST     \\
  392   & CN  +             O$_2$        & $\Longrightarrow$        & CO  +         NO           & 5.12e-12 & -0.49   & -5.2     & UMIST     \\
  416   & NO  +             NO           & $\Longrightarrow$        & N$_2$  +         O$_2$     & 2.51e-11 & 0.00    & 30653.0  & UMIST     \\
  7601  & NH  +             O$_2$        & $\Longrightarrow$        & NO  +         OH           & 2.54e-14 & 1.18    & 312.0    & UMIST     \\
  \hline
        &                                & Collisional Dissociation &                            &          &         &          &           \\
  \hline
  194   & NH  +             NH           & $\Longrightarrow$        & H   +         H   +  N$_2$ & 1.16e-09 & 0.00    & 0.0      & UMIST     \\
  205   & NH  +             NO           & $\Longrightarrow$        & H   +         N$_2$  +  O  & 7.40e-10 & 0.00    & 10540.0  & UMIST     \\
  4060  & H   +             H$_2$        & $\Longrightarrow$        & H   +         H   +  H     & 4.67e-07 & -1.00   & 55000.0  & UMIST     \\
  4061  & CH  +             H            & $\Longrightarrow$        & C   +         H   +  H     & 6.00e-09 & 0.00    & 40200.0  & UMIST     \\
  4062  & H   +             OH           & $\Longrightarrow$        & H   +         H   +  O     & 6.00e-09 & 0.00    & 50900.0  & UMIST     \\
  4067  & H   +             O$_2$        & $\Longrightarrow$        & H   +         O   +  O     & 6.00e-09 & 0.00    & 52300.0  & UMIST     \\
  4069  & H$_2$  +             H$_2$     & $\Longrightarrow$        & H   +         H   +  H$_2$ & 1.00e-08 & 0.00    & 84100.0  & UMIST     \\
  4070  & CH  +             H$_2$        & $\Longrightarrow$        & C   +         H   +  H$_2$ & 6.00e-09 & 0.00    & 40200.0  & UMIST     \\
  4071  & H$_2$  +             OH        & $\Longrightarrow$        & H   +         H$_2$  +  O  & 6.00e-09 & 0.00    & 50900.0  & UMIST     \\
  4074  & H$_2$  +             O$_2$     & $\Longrightarrow$        & H$_2$  +         O   +  O  & 6.00e-09 & 0.00    & 52300.0  & UMIST     \\
  4076  & CO  +             M            & $\Longrightarrow$        & C   +         M   +  O     & 2.79e-03 & -3.52   & 128700.0 & BDDG76    \\
  7002  & CO  +             H            & $\Longrightarrow$        & C   +         H   +  O     & 2.79e-03 & -3.52   & 128700.0 & BDDG76    \\
  7585  & CH  +             O$_2$        & $\Longrightarrow$        & CO  +         H   +  O     & 1.14e-11 & 0.00    & 0.0      & UMIST     \\
  \hline
  \label{table:cno_rn}
\end{longtable}
\twocolumn

The carbon enhancement phenomenon is represented by a number of molecular carbon
features, including the strong CH G-band feature at $\wlAA{4300}$
\citep{grayStellarSpectral2009}, the \CT\ feature at $\wlAA{5636}$
\citep{greenInnocentBystanders2013},
the Swan bands (\CT) at $\wlAA{5635}$ and $\wlAA{5585}$, and the $\wlAA{3883}$
CN band \citep{harmerStellarCN1973}. \citet{kochPurveyorsFine2019} also used CN
features to identify CN-strong and CN-weak stars in globular clusters. Overall,
the spectral synthesis in cool carbon stars from $4000-10000~\rAA$ shows that the
region harbours many CH, CN, and \CT\ lines.

CO has a very high bond-dissociation energy of $11.16$~eV
\citep{marchAdvancedOrganic2001}. It
is the key stable state within the chemical network. In the regions in which
molecular features form, it is energetically favourable to form CO
over CH and OH, for instance. CO therefore dictates the relative yield of other carbon-
and oxygen-bearing molecules. Generally, C and O will be largely consumed to
form CO, and any excess then forms other molecular species. With a C/O ratio
lower than 1 (e.g. for solar composition) and at temperatures that allow for
molecular formation, most of the carbon is locked into CO,
leaving very little to form other carbonic molecules. With a C/O ratio greater
than 1 (e.g. certain carbon-enhanced stars), oxygen is instead used
up first, and more carbonic molecules form (see Fig.~\ref{fig:cemp_comparison}).

We included OH to investigate the effect of the C/O ratio on molecular species.
OH provides an important symmetry to CH when considering the evolution of C, O,
CH, OH, and CO \citep{gallagherIndepthSpectroscopic2016a, gallagherIndepthSpectroscopic2017}.
As the number of non-CO carbon-bearing molecules heavily depends
on the C/O ratio, so too does the evolution of OH.
\subsection{Numerical method}
\label{subsec:numerical_method}
We used \cobold, a conservative finite-volume hydrodynamics solver capable
of modelling surface convection, waves, shocks, and other phenomena in
stellar objects \citep{freytagSimulationsStellar2012}.
The hydrodynamics, radiation transfer, and chemistry were treated via operator
splitting and were solved on a Cartesian grid in a time-dependent manner.
The chemistry was
solved after the hydrodynamics and radiative transfer time steps. Standard
directional splitting along the directions of the 1D operators was used. A Roe
solver computed all updates in a single step, where higher-order terms in time
are provided based on the applied reconstruction scheme.

Radiative transfer was solved frequency dependently (non-grey) under the LTE
assumption using a multiple short-scale
characteristic scheme \citep{steffenRadiationTransport2017}. The opacity tables
use 12 frequency bins and are consistent with the atomic abundances used
for the chemistry input. The model does not treat frequency-dependent
photodissociation of chemical species or heating and cooling via reactions.
The equation of state is also consistent with the abundances used in the
chemistry input and assumes the formation of molecules in instantaneous
equilibrium.

All models used in this work were created by taking a thermally
relaxed \cobold\ model output and adding quantity-centred (QUC) quantities
\citep{freytagSimulationsStellar2012,wedemeyer-bohmCarbonMonoxide2005}.
These QUC quantities allow the user to arbitrarily add cell-centred quantities
to the simulation. Here, each QUC quantity stores the number densities of a
single chemical species
across all grid cells. The QUC quantities were advected as prescribed by the
velocity field.
Periodic boundary conditions were implemented on the lateral edges of the
computational domain. The lower boundary layer was open with inflowing entropy
and pressure adjustment, while the top layer was transmitting. The number densities
in ghost cells were copied from the nearest cells in the computational domain,
but were scaled to the mass density of these cells. In this way, the chemistry was
still consistent across the boundary, and the number densities of the elements
were almost perfectly conserved. We only present 3D models in this work as we
focus on the stellar photosphere and it was shown that 1D models are more
insensitive to a change in CNO abundances \citep{plezAnalysisCarbonrich2005,
  gustafssonGridMARCS2008, masseronStellarNucleosynthesis2008}

The output of the model atmosphere was stored in a sequence of recorded flow
properties, commonly called a sequence of snapshots. Each snapshot also
functioned as a start model to restart a simulation, or as a template to start a
new simulation. To compare the time-dependent
chemistry, the same reaction network was solved on a background static snapshot
(i.e. a single snapshot without taking advection into account) until the
chemistry reached a steady state. This is similar to the treatment of chemistry
in equilibrium, but in this case, we still solved the kinetic system instead of
relying on equilibrium constants. The method for solving the chemistry
independently of the hydrodynamics in post-processing is described in
Sec.~\ref{subsec:gcrn}.

We used $\text{five}$ different 3D model atmospheres that differed in metallicity and
chemical composition. A description of the model parameters is given in
Table~\ref{table:model_overview}. Each model had an effective temperature of
6250 K, a surface gravity of $\log(g) = 4.00$, a resolution of
$140 \times 140 \times 150$ cells, and an extent of $26 \times 26 \times 12.7$
Mm ($x \times y \times z$).
We used standard abundances from the CIFIST grid \citep{caffauSolarChemical2011}
and initialised the molecular species to a number density of $10^{-20}$
g cm$^{-3}$. The models are referred to in the rest of their work by their ID,
namely AM1, AM2, AM3, AC1, and AC2.
The models in this study did not use the MHD module and hence represent only
quiet stellar atmospheres without magnetic fields.

\begin{table}[h!]
  \begin{center}
    \scalebox{0.87}{
      \begin{tabular}{ l r c c r r r l}
        \toprule
        Model ID & [Fe/H]  & A(C)   & A(O)   & $\log$ C / O & Internal ID   \\
        \midrule
                 &         &        &        &              & (d3t63g40)    \\
        \midrule
        AM1      & $+0.00$ & $8.41$ & $8.66$ & $-0.25$      & mm00          \\
        AM2      & $-2.00$ & $6.41$ & $7.06$ & $-0.65$      & mm20          \\
        AM3      & $-3.00$ & $5.41$ & $6.06$ & $-0.65$      & mm30          \\
        \midrule
        AC1      & $-3.00$ & $7.39$ & $7.66$ & $-0.27$      & mm30c20n20o20 \\
        AC2      & $-3.00$ & $7.39$ & $6.06$ & $+1.33$      & mm30c20n20o04 \\
        \midrule
        \bottomrule
      \end{tabular}
    }
  \end{center}
  \caption{Model atmosphere parameters for the five models used in the study.
    Each model has $\Teff = 6250$ K and $\log{g} = 4.00$, a resolution of
    $140\times140\times150$ cells, and an extent of $26\times26\times12.7$ Mm
    ($x\,\times\,y\,\times\,z$). The abundances
    for each model are consistent with those in the respective opacity tables,
    and we use the internal ID to refer to each model uniquely within this
    work.}
  \label{table:model_overview}
\end{table}
\subsection{Time-dependent chemistry}
\label{subsec:method_td_chem}

The radiation-(magneto)hydrodynamics code \cobold\ includes a time-dependent
chemical kinetics solver \citep{wedemeyer-bohmCarbonMonoxide2005,
  freytagSimulationsStellar2012} that has so far been used to
investigate the solar photosphere and chromosphere in two and three dimensions.
The code includes modules to advect passive tracers and to solve a chemical
reaction network using these passive tracers. Generally, these passive tracers are added to an already
thermally relaxed model atmosphere in
order to initialise a model with a time-dependent chemistry. When it is initialised,
\cobold\ then solves the chemistry for each cell at each time step (alongside
the equations of hydrodynamics and radiation transfer), and the species are
advected as prescribed by the velocity field. That is, the time-dependent
chemistry for a species $n_i$ takes the form
\begin{equation}
  \frac{\partial n_i}{\partial t} + \nabla \cdot (n_i \vec{v}) = S,
\end{equation}
where $\vec{v}$ is the velocity field, and $S$ is a source term. The source term
is given by the rate of formation and destruction of each species characterised
by the reactions in the network.

Each chemical reaction can be written as a differential equation describing the
destruction and formation of species, and together, the reactions form an
ordinary differential equation (ODE) system. We considered all reactions in this
work to follow mass-action kinetics. The rate of a reaction is then given by
\begin{equation}
  \label{eq:mass_action_rate}
  w_r = k \prod_j{n_j}
  ,\end{equation}
where $k$ is the rate coefficient, and the product over $n_j$ includes the
stoichiometry of either the reactants (forward reaction) or products (reverse
reaction).
For a generic reaction $r$ with a forward rate coefficient $k_1$ and reverse rate
coefficient $k_2,$
\begin{equation}
  a\mathrm{A} + b\mathrm{B} \rightleftharpoons ^{k_1}_{k_2} c\mathrm{C} + d\mathrm{D},
\end{equation}
the rates of change of the generic species A, B, C, and D in the reaction $r$ are
related via
\begin{equation}
  -\frac{1}{a}\left(\frac{\partial n_\mathrm{A}}{\partial t}\right)_r = -\frac{1}{b}\left(\frac{\partial n_\mathrm{B}}{\partial t}\right)_r = \frac{1}{c}\left(\frac{\partial n_\mathrm{C}}{\partial t}\right)_r = \frac{1}{d}\left(\frac{\partial n_\mathrm{D}}{\partial t}\right)_r.
\end{equation}
Eq.~(\ref{eq:mass_action_rate}) then gives the forward and reverse reaction
rates $w_1$ and $w_2$ as
\begin{equation*}
  w_1 = k_1 n_\mathrm{A}^a n_\mathrm{B}^b, ~~~
  w_2 = k_2 n_\mathrm{C}^c n_\mathrm{D,}^d
\end{equation*}
respectively. We can then construct the differential
$\left(\frac{\partial{n_i}}{\partial t}\right)_r$
for a species $n_i$ and reaction $r$. The full time-dependent chemical evolution
of species $n_i$ is then given by the sum over the reactions $r$,
\begin{equation}
  \frac{\partial n_i}{\partial{t}} = \sum_r \left(\frac{\partial n_i}{\partial t}\right)_r.
  \label{eq:chem_rxn_sys}
\end{equation}
Particle conservation is ensured due to the overall mass continuity
equation and the stoichiometry of each chemical reaction. Because only neutral
species are considered, no explicit charge conservation is included.

Due to the
high computational expense of computing time-dependent chemistry across a large
grid, parallelisation is highly recommended. Along with the increased
memory load of storing the number densities of QUC species, this limits the size of
the network that can be treated time dependently. Even with these steps, solving
the chemistry is still the most time-intensive step, taking upwards of $75$\% of
the total runtime.

The DVODE solver \citep{hindmarshSUNDIALSSuite2005} was used to solve the system
of chemical
kinetic equations, making use of the implicit backward differentiation formula
(BDF). The solver uses an internally adjusted adaptive time step, which is  a
requirement when we consider that the system of equations is often very stiff.
The solution of the final number densities is provided after the full
hydrodynamics time step.

For stability, we used first-order reconstruction schemes for both the
hydrodynamics and advection of QUC quantities. Higher-order schemes were found
to cause some grid cells to extrapolate beyond the equation-of-state tables or
low number densities to become negative. This was not a consistently
reproducible effect for a given grid cell, meaning that its source could lie in
single-precision numerical errors.
\subsection{Rate coefficients}
\label{subsec:rate_coefficients}

The rate coefficient (sometimes rate constant) of a chemical reaction is an
often empirically determined quantity dependent on the temperature. Many of the
reactions presented in this work are unfortunately defined outside of their
temperature limits
simply because we lack studies of chemical reactions in high-temperature
regions such as stellar photospheric layers. An uncertainty
is also associated with the rate coefficients themselves. Despite these shortcomings, the chosen reaction rates are thought to describe the evolution of our species
reasonably well.

The Arrhenius rate coefficient is commonly written as

\begin{equation}
  k = \alpha \exp\left(-\frac{E_a}{RT}\right),
\end{equation}
where $\alpha$ is a pre-exponential factor representing the fraction of species
that would react if the activation energy $E_a$ were zero, $R$ is the gas
constant, and $T$ is the temperature. We used the modified Arrhenius equation, which explicitly includes the temperature dependence

\begin{equation}
  k = \alpha \left(\frac{T}{300 ~[\mathrm{K]}}\right)^\beta \exp\left(-\frac{\gamma}{T}\right),
\end{equation}
where $\beta$ is a fitted constant for a given reaction, and
$\gamma = \frac{E_a}{R}$ characterises the activation energy. For a reversible
reaction, the forward and reverse coefficients are related to the dimensional
equilibrium constant $K'_\mathrm{eq}$ by

\begin{equation}
  K'_\mathrm{eq} = \frac{k_f}{k_r},
\end{equation}
where $k_f$ and $k_r$ are the forward and reverse rate coefficients,
respectively. This equilibrium constant can be used to determine the chemical
equilibrium of a given composition, defined when all forward and reverse
processes are balanced \citep{blecicTEACODE2016,stockFastChemComputer2018}.
As our reaction network contains irreversible reactions in the thermodynamic
domain under study, equilibrium constants cannot be determined for
each chemical pathway. Hence, we studied the equilibrium chemistry by solving
the chemical kinetics until the chemistry reached a steady state. In the
absence of processes such as advection, this steady state should correspond to
chemical equilibrium.
\subsection{Steady-state chemistry}
\label{subsec:gcrn}

Steady-state chemistry was treated by solving the
chemical kinetic system on a background model atmosphere (a single static
snapshot), neglecting advection.
The chemistry was evolved long enough to reach a steady state in which processes were
balanced for each grid cell. The formulation of
the final system of equations is the same as that in Eq \ref{eq:chem_rxn_sys}.
In this way, we were able to evaluate the time-dependent effects of advection
when compared to the statically post-processed chemistry in steady state.

To solve the chemistry on a background \cobold\ model snapshot, we present the code called
graph chemical reaction network (GCRN)\footnote{\url{https://github.com/SiddhantDeshmukh/graphCRNs}}. GCRN strictly handles a chemical kinetics problem and is able to evaluate the solution at
arbitrary times, provided the chemical network, initial number densities, and
temperature. The chemistry is solved isothermally in each cell. GCRN is able to
read and
write chemical network files in the format required by \cobold\ as well as that
of KROME \citep{grassiKROMEPackage2014}. The code is written in Python and primarily relies on the numpy, scipy, and networkx libraries.

The numerical solver is the same as wasused in the time-dependent case, namely
DVODE with the BDF method. By default, the absolute tolerance was set to $10^{-30}$
and the relative tolerance to $10^{-4}$.
The Jacobian was computed and evaluated within the DVODE solver itself, but GCRN
supports a user-supplied Jacobian matrix. GCRN can also automatically compute an
analytical Jacobian based on the equation system and pass this to the solver.
Supplying a Jacobian to the solver can help improve stability, but it was not
necessary in this work.

GCRN first represents the system of chemical reactions as a weighted, directed
graph (see e.g. \citet{vanderschaftComplexDetailed2015,hornNecessarySufficient1972}).
The vertices of the
graph are the left and right sides of the chemical reactions, hereafter
complexes, while the edges represent the reactions themselves. The weights of
the edges are the reaction rates, evaluated for the provided temperature and
initial number densities. For a reaction network with $c$ complexes and $r$
reactions, its directed (multi)graph\footnote{allows for multiple edges between
  vertices} $G$ can be characterised by its $c \times r$ incidence matrix
$\matr{D}$, which represents the connection between vertices and edges, that is,
which edges connect which vertices. Each column of $\matr{D}$ corresponds to an
edge (a reaction) of $G$. The ($i,j$)th element of $\matr{D}$ represents the
reaction $j$ containing complex $i$. It is $+1$ if $i$ is a
product, and $-1$ if $i$ is a reactant. For $s$ species, the $s \times c$
complex composition matrix $\matr{Z}$ describes the mapping from the space of
complexes to that of species, that is, it describes which species make up which
complexes. Multiplying $\matr{Z}$ and $\matr{D}$ yields the $s \times r$
stoichiometric matrix $\matr{S} = \matr{Z}\matr{D}$. Finally, to include the
mass-action kinetics, we required a vector of reaction rates $\vec{v}(\vec{x})$
as a function of the species vector $\vec{x}$. In general, for a single reaction
with a reactant complex $C$ specified by its corresponding column
$\vec{z}_C = [z_{C, 1} \dots z_{C, s}]^T$ of $\matr{Z}$, the mass
action kinetic rate with the rate coefficient $k$ is given by

\begin{equation}
  kx_1^{z_{C, 1}}x_2^{z_{C, 2}} \dots x_m^{z_{C, s}},
\end{equation}

or more concisely,

\begin{equation}
  k\exp(\vec{z}_C^T\mathrm{\bf{ln}}(\vec{x})), 
\end{equation}

where $\mathrm{\bf{ln}(\vec{x})}$ is defined as an element-wise operation producing
the vector $[\ln(x_1) \dots \ln(x_s)]^T$. Similarly, the element-wise operation
$\mathrm{\bf{exp}}(\vec{y})$ produces the vector
$[\exp({y_1}) \dots \exp({y_s})]^T$. With
this, the mass-action reaction rates of the total network are given by

\begin{equation}
  v_j(\vec{x}) = k_j \exp\left(\vec{z}_{j}^T\mathrm{\bf{ln}}(\vec{x})\right), j = 1, \dots,r
  .\end{equation}

This can be written compactly in matrix form. We defined the $r \times c$ matrix
$\matr{K}$ as the matrix whose ($j,\sigma$)th element is the rate coefficient
$k_j$ if the $\sigma$th complex is the reactant complex of the $j$th reaction,
and zero otherwise. Then,

\begin{equation}
  \vec{v}(\vec{x}) = \matr{K}\,\mathrm{\bf{exp}}\left(\matr{Z}^T\mathrm{Ln}(\vec{x})\right)
  ,\end{equation}

and the mass-action reaction kinetic system can be written as

\begin{equation}
  \dot{\vec{x}} = \matr{Z}\,\matr{D}\,\matr{K}\,\mathrm{\bf{exp}}\left(\matr{Z}^T\mathrm{\bf{ln}}\vec{x}\right)
  .\end{equation}

The formulation is equivalent to that in Eq.~(\ref{eq:chem_rxn_sys}), with the
stoichiometric matrix $\matr{S} = \matr{Z}\,\matr{D}$ supplying the stoichiometric
coefficients and the rate vector
$\vec{v}(\vec{x}) = \matr{K}\,\mathrm{\bf{exp}}\left(\matr{Z}^T\mathrm{\bf{ln}}(\vec{x})\right)$
supplying the mass-action kinetic rates. A detailed explanation on the graph
theoretical formulation and further analyses can be found in
\citet{vanderschaftNetworkDynamics2016}.

A graph theoretical approach allows us to investigate certain behaviours across
chemical pathways, such as the timescales of processes and the importance of
certain species. These graph representations are only created and
accessed upon request and are not used when solving the kinetic system. The
graph representations then allow for the analysis of the network in more depth
before and after solving the system. In solving the actual kinetic system, only
the rates vector $\vec{v}(\vec{x})$ changes based on the change in number
densities. We refer to Sec~\ref{subsec:timescales-pathways} for an in-depth analysis of
the representation of the CRN as a graph.

A drawback of the Python version of GCRN is its low efficiency compared to
compiled languages. Although
we have implemented a few optimisations, computing the chemical evolution for
many snapshots in 3D is still computationally challenging. We therefore used the Julia library Catalyst.jl
\footnote{\url{https://catalyst.sciml.ai/dev/}} for
steady-state calculations across many 3D snapshots. GCRN was used primarily to
evaluate 2D slices, 1D averages, and timescales, while large 3D steady-state
calculations were performed in Julia. The results are identical between the two.
\section{Results}
\label{sec:results}

\subsection{Time-dependent versus steady-state chemistry}

\begin{figure*}
  \centering
  \begin{subfigure}{0.33\textwidth}
    \includegraphics[width=\textwidth]{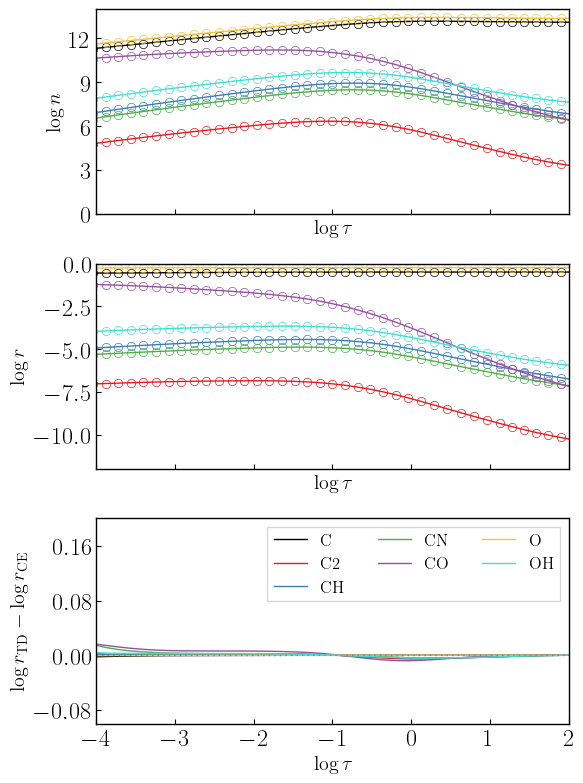}
    \caption{}
    \label{subfig:mm00_tau_avg}
  \end{subfigure}
  \begin{subfigure}{0.33\textwidth}
    \includegraphics[width=\textwidth]{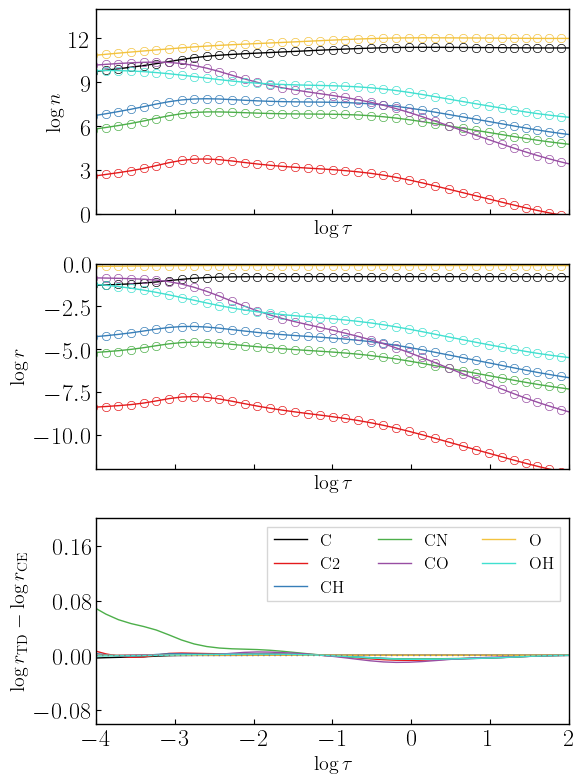}
    \caption{}
    \label{subfig:mm20_tau_avg}
  \end{subfigure}
  \begin{subfigure}{0.33\textwidth}
    \includegraphics[width=\textwidth]{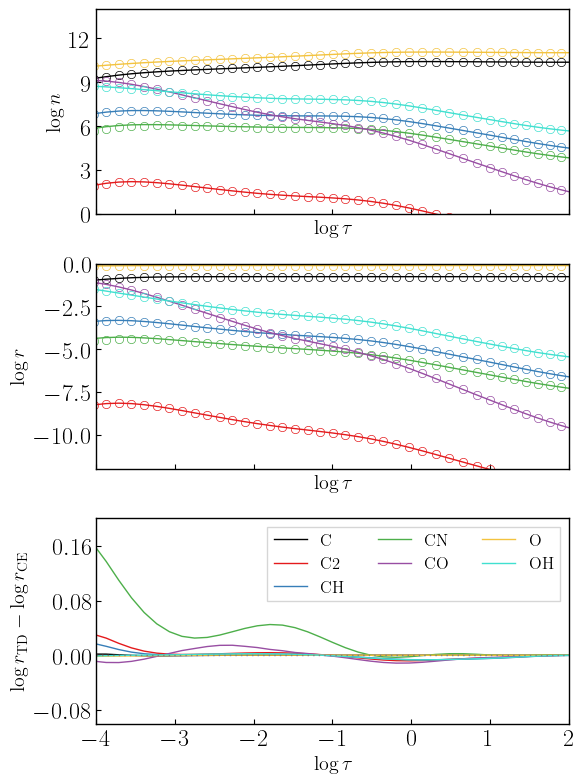}
    \caption{}
    \label{subfig:mm30_tau_avg}
  \end{subfigure}
  \caption{Mixing ratios and deviations from chemical equilibrium for the AM1,
    AM2, and AM3 models. In the first and second panels, solid lines show the
    time-dependent quantities, and the hollow points show the equilibrium
    quantities.
    \textbf{Left.} [Fe/H]~=~$0.0$. \textbf{Centre.} [Fe/H]~=~$-2.0$.
    \textbf{Right.} [Fe/H]~=~$-3.0$.}
  \label{fig:metallicity_tau}
\end{figure*}

We investigated the results of time-dependent (TD) chemistry compared to
equilibrium (Eqm) chemistry in 3D. For all models, chemical
equilibrium is generally held below $\log{\tau} = 1$.
Fig.~\ref{fig:metallicity_tau} shows the absolute number densities and
mixing ratios of species across the photosphere for both the time-dependent and
steady-state chemistry. The mixing ratio is defined as
\begin{equation}
  r = \frac{n_i}{n_\mathrm{total}},
\end{equation}
where $n_i$ is the number density of species $i,$ and $n_\mathrm{total}$ is the
number density of all species excluding H, H$_2$ , and M. In this way, the mixing
ratio describes the relative abundances of important atomic and molecular
species in a given volume. H, H$_2$ , and M are much more abundant than other
species, and including these species simply scales the relevant quantities down.
We characterise deviations by considering the ratio of TD to CE mixing ratios,
which is equivalent to considering the ratio of TD to CE number densities.

Molecular chemistry is clearly in equilibrium in the deeper photospheric layers,
generally below $\log{\tau} = 1$. This is expected because the high temperatures in
this collision-dominated regime result in very short timescales (much shorter
than characteristic hydrodynamical timescales). In essence, the assumption of
chemical equilibrium holds in these regimes. Significant deviations are not
present in the AM1 model, but appear above $\log{\tau} \approx -2$ in model AM2
and above $\log{\tau} \approx -1$ in model AM3.
In all cases in which the deviations are non-zero, the time-dependent
chemistry is affected by hydrodynamics such that there is insufficient time to
reach a local chemical equilibrium.

As expected, a decreasing metallicity decreases the number of molecular species
that can be formed. The deviations from equilibrium molecular number densities
increase with decreasing metallicity because the chemical timescales are slower. The
largest deviations are seen in \CT\ and CN in model AM3, where they reach up to
$0.15$ dex at $\log{\tau} = -4$. The deviations for the other molecules similarly increase
with increasing height. These positive deviations are balanced by (smaller)
negative deviations in CO. Essentially, there is insufficient time to form the
equilibrium yield of CO in these thermodynamic conditions, and the yield
of species that would react to form CO is therefore higher.

\begin{figure*}
  \centering
  \scalebox{0.65}{
    \includegraphics[]{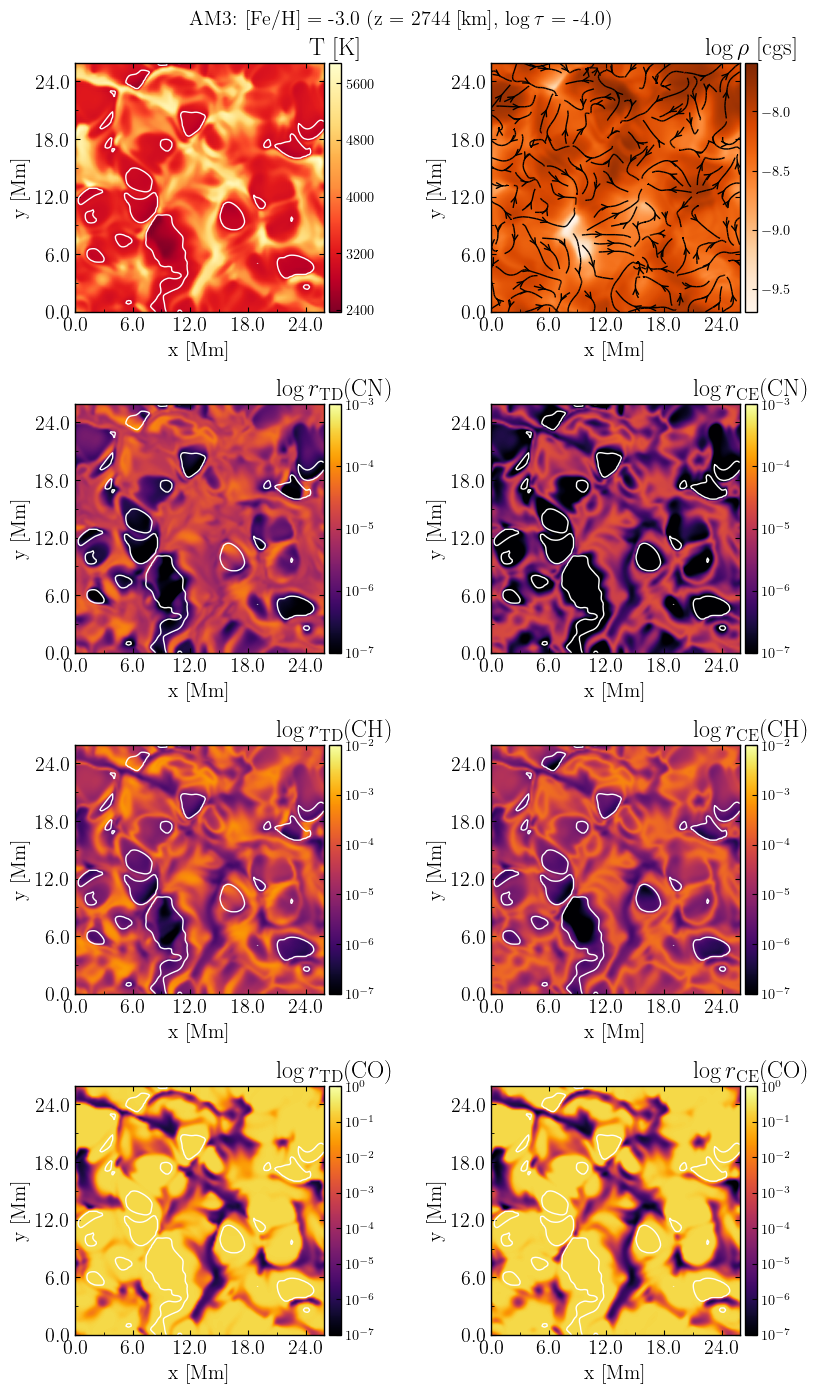}
  }
  \caption{Mixing ratios of molecular species in a horizontal slice through
    the photosphere in the AM3 model. \textbf{Left.} Time dependent.
    \textbf{Right.} Equilibrium. Molecular formation follows a reversed
    granulation pattern. The effect of finite chemical
    timescales is most prominent when contrasting warm and cool regions in CN
    and CH; CO is seen to be relatively close to CE, as confirmed by
    Fig.~\ref{subfig:mm30_tau_avg} at $\log \tau = -4$. The white contour
    traces a temperature of $4500$~K.}
  \label{fig:mm30_xy}
\end{figure*}

\begin{figure*}
  \centering
  \scalebox{0.62}{
    \includegraphics[]{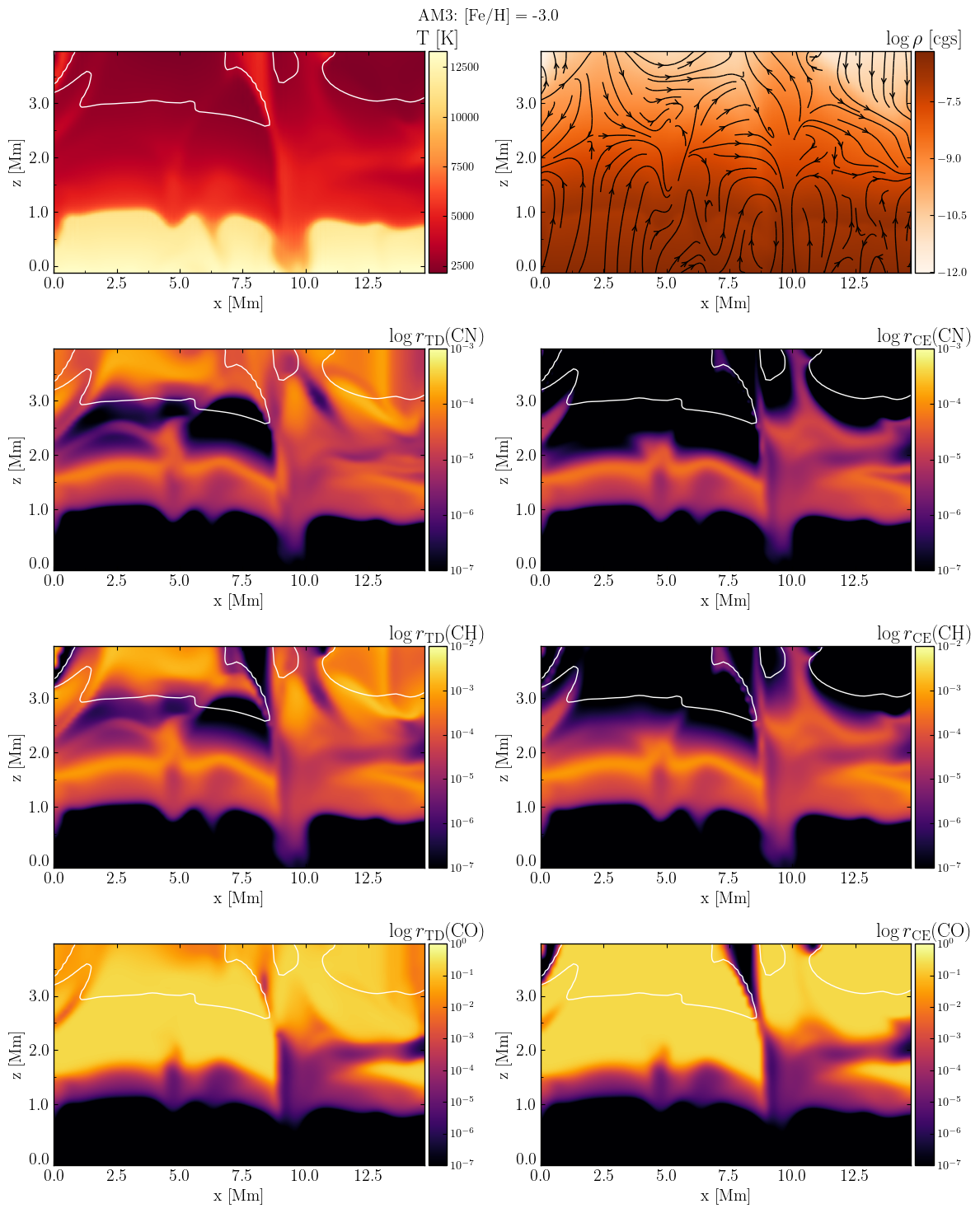}
  }
  \caption{
    Time-dependent mixing ratios of molecular species in a vertical slice
    through the photosphere above $\log {\tau} = 1$ in the AM3 model atmosphere.
    The white contour traces a temperature of $4500$~K.
    The colour scale is the same for all molecular species.
  }
  \label{fig:mm30_xz}
\end{figure*}

Differences that are often present around local features such as shocks can be lost in the global picture (averaging over space and time). Even though the
chemistry is mostly in equilibrium throughout the atmosphere, investigating
cases in which it is out of equilibrium can lead to an understanding of the
hydrodynamical effects as well as to insights into where approximations of chemical
equilibrium break down.
Figs.~\ref{fig:mm30_xy} and \ref{fig:mm30_xz} show the time-dependent mixing
ratios in a horizontal and vertical slice through the AM3 model atmosphere,
respectively.

Fig.~\ref{fig:mm30_xy} shows deviations from CE in CN in and around cool
features. Mass-action rates depend on temperature, and therefore, cooler cells lead to
longer chemical timescales. The instantaneous CE therefore predicts faster
dissociation than is possible within the time-dependent scheme. In higher
layers, the same reasoning applies, leading to positive deviations in CN, CH,
and \CT\ , offset by negative deviations in CO.

The vertical slice in Fig.~\ref{fig:mm30_xz} shows the evolution of chemistry in
various layers and highlights a shock in the upper photosphere. Deviations from
CE are seen in all species in higher layers, with the shock being the most
prominent example. In CE, all molecular species are immediately dissociated,
while the time-dependent shows that even in these higher-temperature regions,
CO is not so quickly depleted. While it may seem counter-intuitive that CO then
shows a small negative deviation from CE, the mean amount of CO in the
time-dependent case is smaller than that in CE. This is reflected in the positive
deviations from CE seen in CH and CN, which, due to mass conservation, are
offset by the negative deviation in CO. Additionally, the reverse trend is also
true, in that the formation of CO after a shock passes is slower than predicted
in CE.

\subsection{Carbon enhancement}
\label{subsec:carbon_enhancement}

For the models presented thus far, oxygen has been more abundant than carbon.
CO, being extremely stable, often dominates the molecular species when it can
form. It is possible, however, that this preference towards CO formation is
influenced by the enhancement of oxygen relative to carbon present in the
atmosphere. We investigated two cases of carbon enhancement in a model
atmosphere with metallicity [Fe/H] = $-3.0$. The first increased both C and O by
$2.0$ dex (AC1), while the second only increased C by $2.0$ dex (AC2).
Nitrogen was also increased by $2.0$ dex. The increase for all elements
included the $0.4$ dex enhancement for alpha elements.

\begin{figure*}
  \centering
  \begin{subfigure}{0.49\textwidth}
    \includegraphics[width=\textwidth]{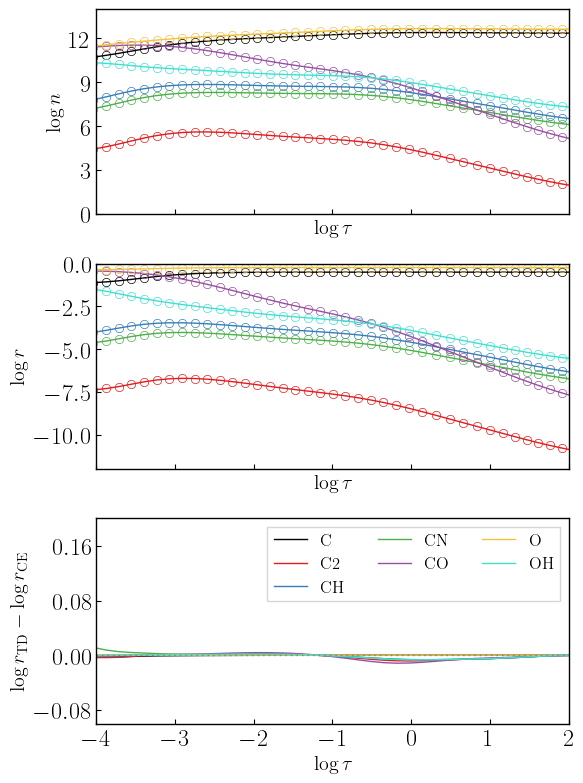}
    \caption{}
    \label{subfig:mm30c20o20_tau_avg}
  \end{subfigure}
  \begin{subfigure}{0.49\textwidth}
    \includegraphics[width=\textwidth]{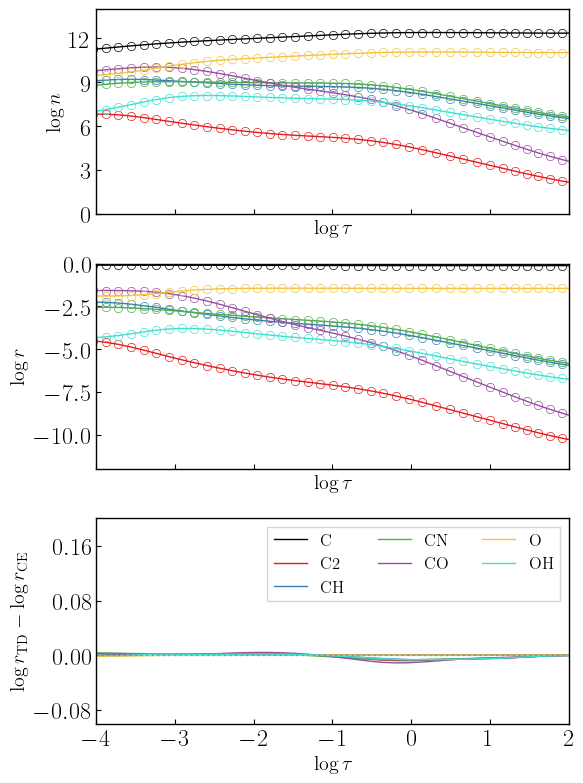}
    \caption{}
    \label{subfig:mm30c20o04_tau_avg}
  \end{subfigure}
  \caption{Mixing ratios and deviations from chemical equilibrium at
      [Fe/H] = $-3.0$ and a carbon enhancement of $+2.0$ dex for two atmospheres
    with different C/O ratios. In the first and second panels, solid lines show
    the time-dependent quantities, and the hollow points show the equilibrium
    quantities.
    \textbf{Left.} Model AC1, C/O = $-0.26$.
    \textbf{Right.} Model AC2, C/O = $+1.33$.}
  \label{fig:cemp_tau}
\end{figure*}

Fig. \ref{fig:cemp_tau} shows the mixing ratios and deviations from equilibrium
for the two CEMP model atmospheres presented in this work.
In model AC1 ($\log$ (C/O) = $-0.26$), more CO and CH is formed than in
the standard metal-poor case, but OH is still more abundant than CH. Almost all
C is locked up into CO, hence the next most-abundant molecular species is OH.
This is analogous to models AM2 and AM3 because O is still more abundant than C.
Carbon-bearing molecules are more abundant than in AM3, but the mixing ratios of CH to OH, for example, clearly show that the carbon
enhancement does not necessarily lead to a large increase in all carbon-bearing
molecular abundances. In model AC2 (C/O = $+1.33$), CO is still the most
abundant species, while CH is more abundant than OH. We observe the opposite
effect compared to models AM2, AM3, and AC1, ib which instead O is locked
up into CO. This results in a significant depletion of OH compared to model
AM3 because there is relatively little O left to form OH because C is overabundant. The depletion of O hinders the formation of further CO, and the chemical
equilibrium is such that atomic C is the most abundant species. All models hence
reinforce the notion that CO is the most stable molecular state in the chemical
network.

Oxygen-bearing species seem to be further out of equilibrium in model
AC1, while carbon-bearing species are further out of equilibrium in model AC2.
Interestingly, deviations from equilibrium decrease in model AC2, in which the C/O
ratio of $+1.33$ means that carbon is more abundant than oxygen. While this favours
the formation of carbon-bearing species such as \CT\ and CH, the formation of CO is
hindered compared to model AC1 by the lack of OH formation, reinforcing the
idea that the pathway for CO formation involving OH is important. The
significantly smaller deviations in model AC2 might suggest that oxygen-bearing
molecules might show larger deviations from chemical equilibrium due to
hydrodynamical effects. All in all, CEMP atmospheres do not seem to be largely
out of chemical equilibrium for the species presented in this work.

\section{Discussion}
\label{sec:discussion}

\subsection{Effects of convection}
\label{subsec:effects_convection}

As material is transported from hotter, deeper photospheric layers to cooler,
higher layes, the conditions for chemistry to equilibriate change. It is
feasible, then, that material from a lower layer can be carried upwards, reach a
new equilibrium state, and later return to a deeper layer. In this process,
molecular species will be present in greater numbers in cooler regions than in hotter regions. If chemistry does not equilibriate faster
than advection occurs, we observe deviations from chemical equilibrium
throughout convection cells. This effect is seen in Fig.~\ref{fig:mm30_xy} for
CN and CH, where features are traced much more sharply in the
equilibrium case than in the time-dependent one. The finite chemical
timescales are responsible for the differences in formation in cool regions, and
dissociation in hot ones. In this layer, the chemical equilibrium approximation
still holds well for CO.
\subsection{Behaviour around shocks}
\label{subsec:shock_behaviour}

\begin{figure*}
  \centering
  \scalebox{0.75}{
    \includegraphics[]{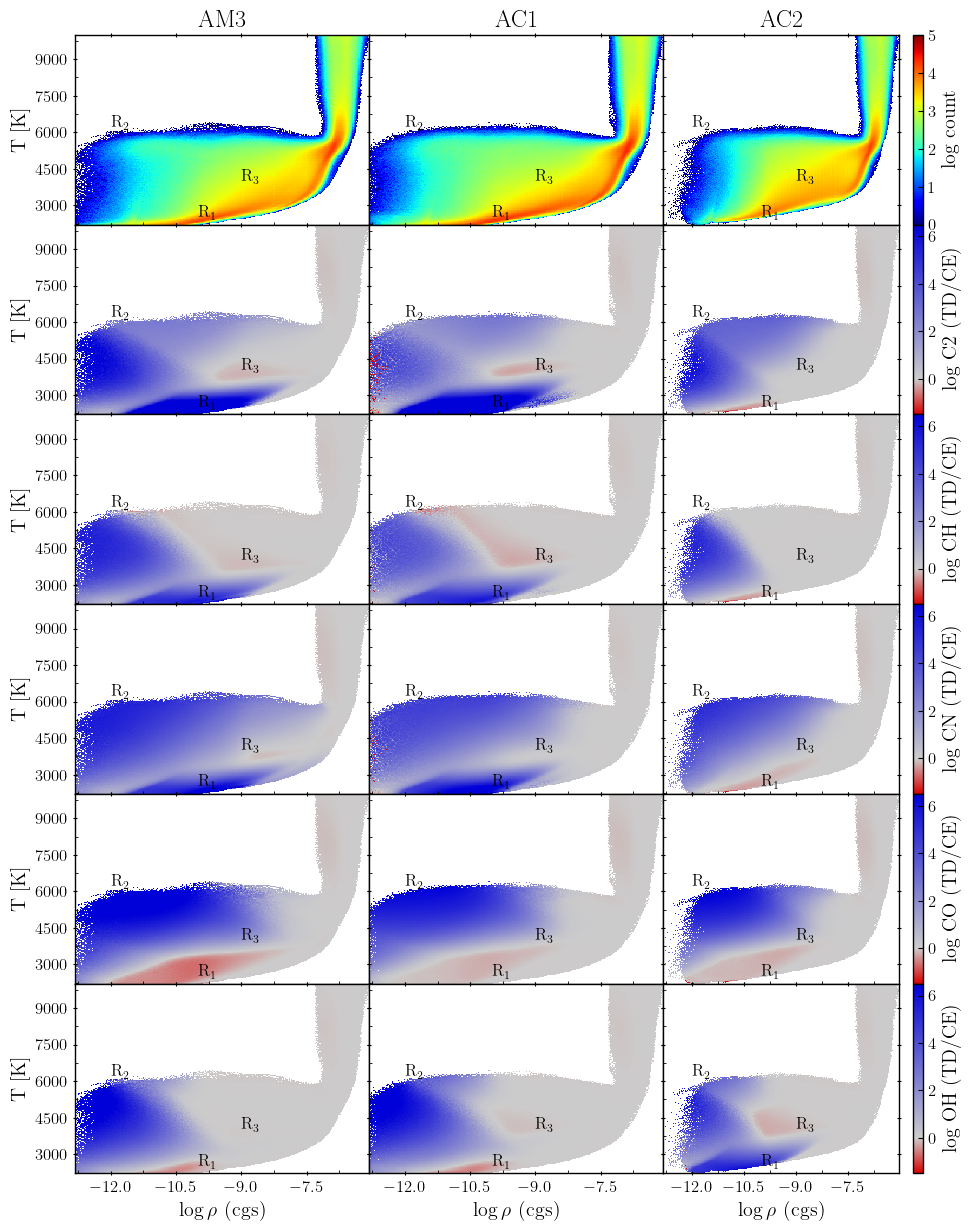}
  }
  \caption{Heat maps of binned quantities for models AM3, AC1, and AC2. Each
    quantity was binned using $20$ snapshots of each 3D model. Deviations from
    equilibrium are seen in three distinct regions, labelled R$_1$
    (convective cells in the upper photosphere), R$_2$ (shock fronts in the
    chromosphere), and R$_3$ (wake of the shock).}
  \label{fig:vaex}
\end{figure*}

While the overall differences in time-dependent and steady-state chemistry are
small when averaged over time and space (horizontally), there can be significant
differences in individual instances in time. In addition to the
shock seen in Fig.~\ref{fig:mm30_xz}, Fig.~\ref{fig:vaex}
shows the deviations from equilibrium molecular chemistry in the photospheres
of the AM3, AC1, and AC2 models. This histogram shows deviations from CE binned
in gas density and temperature across all $20$ snapshots. The top panel gives
the bin counts, showing the difference between background material (high density
of points) and transient states (low density of points).

Although the background material is generally in
equilibrium, three interesting regimes emerge where the molecular chemistry is
clearly out of equilibrium, labelled R$_1$, R$_2,$ and R$_3$. R$_1$ is the
regime of convection in the upper photosphere and chromosphere,
where hot material is advected upwards to a new layer faster than the molecular
chemistry can reach equilibrium. When this material cools and falls, it can
sometimes reach very high velocities (around $10$~\kms) exceeding the local sound
speed. This supersonic material of the shock front is captured in the regime
R$_2$. Equilibrium chemistry predicts an almost instantaneous dissociation of
molecular species, while the time-dependent case models this over a finite
timescale. An excess of molecular species is therefore present in the time-dependent
case. Finally, the regime R$_3$ is the wake of the shock, where material has
cooled and is subsonic. The slower chemical timescales in this regime lead to a
depletion of molecular species in the time-dependent case. CO is an outlier
here; it is still present in slight excess in R$_3$ as it does not dissociate as
quickly as the other molecular species in the shock.

Models AC1 and AC2 show opposite trends in regimes R$_1$ when considering CH,
CN, \CT\ , and OH. In model AC1 ($\log \mathrm{C/O} = -0.26$), the carbon-bearing
molecules are more abundant in the time-dependent case, and OH is depleted.
Model AC2 ($\log \mathrm{C/O} = +1.33$) instead has fewer carbon-bearing molecules, and OH is more abundant. This is due to the relative abundances
of C and O. The chemical timescales depend on the abundances of C and O, so that the
oxygen-rich atmosphere AC1 has slower dissociation rates for carbon-bearing
molecules but a higher yield because the formation rates for OH are faster (and
vice versa for the carbon-rich atmosphere AC2). Since CO is a stable end-product
of most reaction pathways, it is not as strongly affected by this phenomenon.

Overall, the differences
between the time-dependent and steady-state treatments in the photosphere are
small, meaning that the chemistry in convection cells is likely not far from its
equilibrium state. This is especially evident when averaging over space and
time. However, it is possible that the effects would become
stronger in stars on the red giant branch (RGB stars) due to larger scale flows
and M-type dwarfs due to
cooler temperatures, although the latter have smaller velocity fields, meaning
that the effects of advection on the evolution of chemical species are reduced.
\citet{wedemeyer-bohmCarbonMonoxide2005} showed that the need for time-dependent
chemistry becomes
increasingly important in the solar chromosphere due to the higher frequency of
shock waves alongside lower chemical timescales, but that the photosphere of the
Sun was generally in chemical equilibrium for CO. We find the same trend when
considering metal-poor dwarf stars: chemical
equilibrium generally holds for the photospheres of these stars when
we average over space and time, and deviations are largely present in their
chromospheres. This further shows the need to include accurate time-dependent
molecular chemistry when modelling stellar chromospheres.
\subsection{1D analysis}
\label{subsec:profile_1d}

\begin{figure}
  \centering
  \scalebox{0.50}{
    \includegraphics[]{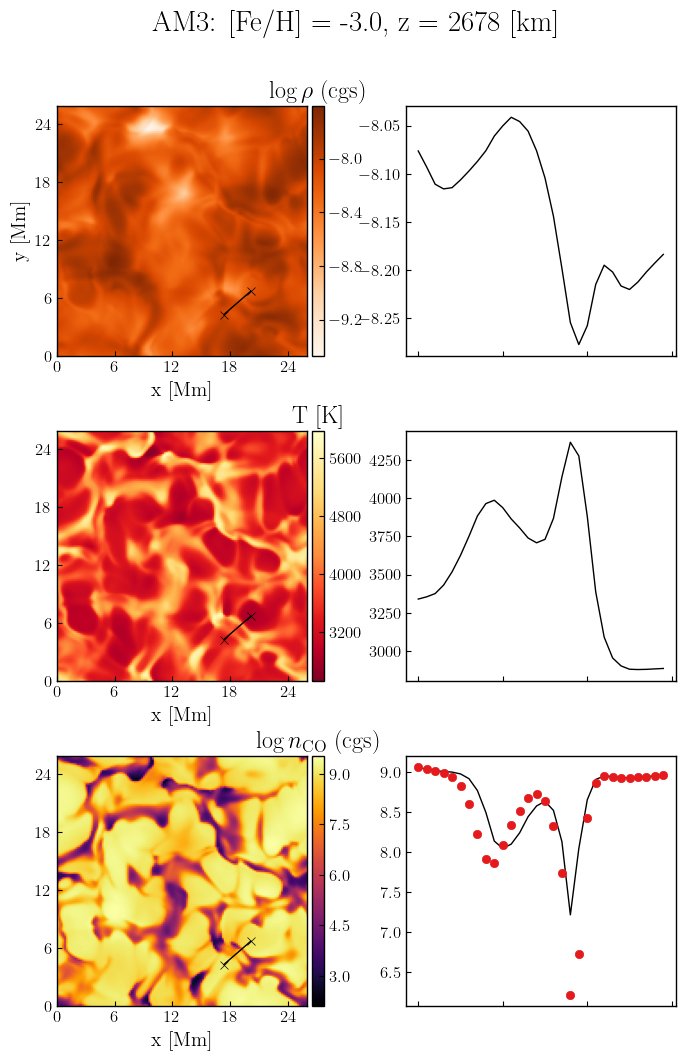}
  }
  \caption{Gas density, temperature, and the number density of CO molecules in a
    slice across the AM3 atmosphere. The left panels show the 2D heat maps of
    these quantities, and the right panels show a 1D cut across a prototypical
    downflow feature, depicted by the
    solid black line in the top panels. The bottom right panel shows
    the time-dependent number density as a solid black line and the equilibrium
    number density as red points.}
  \label{fig:mm30_xy_1d}
\end{figure}

A 1D horizontal cut through the atmosphere shows the instantaneous
variations in the parameters and can help identify patterns. Due to mass-action
kinetics, the chemical timescales depend on the gas density and temperature.
Fig.~\ref{fig:mm30_xy_1d} shows
profiles of these quantities alongside the time-dependent and equilibrium number
densities of CO across a prototypical downflow feature in the chromosphere of model AM3.

The equilibrium CO number density changes much more sharply across the feature
than in the time-dependent case. This shows the finite chemical timescales. The number densities are also more sensitive to fluctuations in
temperature, as seen towards the end, when the gas density changes but
the temperature is constant. The equilibrium number densities show sharp
discontinuities due to the
vastly different chemical timescales around the shock front. While these are
implausible, the average number densities are very similar (as shown in
Fig.~\ref{fig:metallicity_tau}), showing that the
shock here is not disruptive enough for CO chemistry to have a profound impact
overall.
\subsection{Timescales and pathways}
\label{subsec:timescales-pathways}

\begin{figure}
  \centering
  \scalebox{0.75}{
    \includegraphics{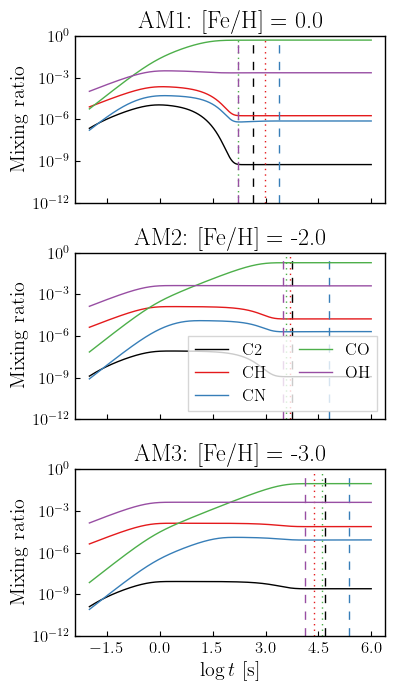}
  }
  \caption{Chemical evolution for $\text{three}$ models at differing metallicities
  at T = $3500$ K, $\rho=10^{-9}$ g cm$^{-3}$ (corresponding to the wake
  of a shock).
  The vertical dash-dotted lines show the time a
  species has to set into equilibrium. A reduction in metallicity leads to
  a corresponding reduction in time to equilibrium and overall yield.}
  \label{fig:metallicity_comparison}
\end{figure}
\begin{figure}
  \centering
  \scalebox{0.75}{
    \includegraphics{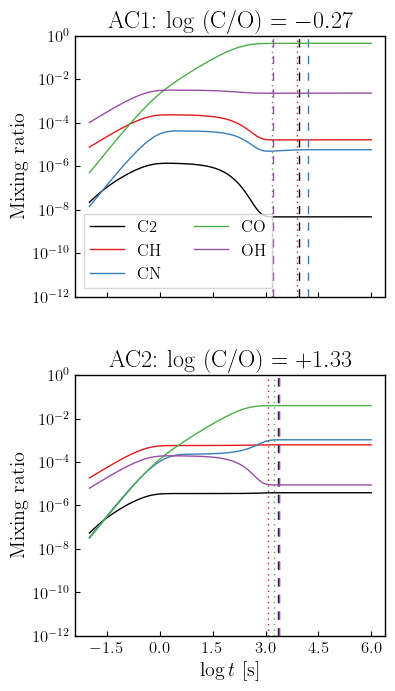}
  }
  \caption{Chemical evolution for $\text{two}$ models with [Fe/H] = $-3.0$, but
  differing C/O ratios at T = $3500$ K, $\rho=10^{-9}$ g cm$^{-3}$
  (corresponding to chromospheric background layers).
  The vertical dash-dotted lines show the time a species has to set into
  equilibrium. In the oxygen-dominated atmosphere, CH is depleted
  compared to OH, while the opposite is true in the carbon-dominated
  atmosphere.}
  \label{fig:cemp_comparison}
\end{figure}

It is perceivable that a metallicity reduction by $2.0$ leads to
timescales that are slower by a factor of $\sim 100$ due to the mass action law.
Additionally, relative abundances have a strong effect, and the overall yield is
lower at lower metallicities. Fig.~\ref{fig:metallicity_comparison} shows the
evolution and equilibrium times for
$\text{three}$ metallicities, and Fig. \ref{fig:cemp_comparison} shows this for the two
CEMP models. The equilibrium times for each model and species of interest are
given in Table \ref{table:metallicity_cemp_times}. Because of the
time-stepping of the solver, these times are not necessarily exact, but they
provide a clear picture of how the species interact. The equilibrium times
here are generally given as the point at which the relative difference in the
number densities falls below a threshold $\epsilon$, that is, $t_\mathrm{eqm}$ is
reached when $\frac{n_{i+1}}{n_i} \leq \epsilon$. We adopted $\epsilon = 10^{-6}$
for this network. Again, because this definition relies on the solver
time-stepping to find $n_i$, $n_{i+1}$, the times are only exact to the times
where the solution is evaluated.

\begin{table}[h!]
  \caption{Time to equilibrium for all models and key molecular species at a
    temperature of $3500$~K and a gas density of $10^{-9}$~g~cm$^{-3}$,
    corresponding to the upper photospheric convection cells. Due to
    the time-stepping of the solver, these times are not exact, but they provide
    a useful picture of how quickly various species set into equilibrium at
    varying chemical compositions.}
  \begin{center}
    \scalebox{0.85}{
      \begin{tabular}{ l r r r r r}
        \toprule
        Model & $t_\mathrm{eqm}$(C$_2$) & $t_\mathrm{eqm}$(CH) & $t_\mathrm{eqm}$(CN) & $t_\mathrm{eqm}$(CO) & $t_\mathrm{eqm}$(OH) \\
              & [s]                     & [s]                  & [s]                  & [s]                  & [s]                  \\
        \midrule
        AM1   & $4.5 \times 10^2$       & $1.0 \times 10^3$    & $2.4 \times 10^3$    & $1.7 \times 10^2$    & $1.7 \times 10^2$    \\
        AM2   & $5.7 \times 10^3$       & $5.1 \times 10^3$    & $6.3 \times 10^4$    & $3.9 \times 10^3$    & $3.2 \times 10^3$    \\
        AM3   & $4.9 \times 10^4$       & $2.4 \times 10^4$    & $2.4 \times 10^5$    & $4.0 \times 10^4$    & $1.3 \times 10^4$    \\
        \midrule
        AC1   & $9.0 \times 10^3$       & $7.7 \times 10^3$    & $1.6 \times 10^4$    & $1.6 \times 10^3$    & $1.6 \times 10^3$    \\
        AC2   & $2.2 \times 10^3$       & $1.2 \times 10^3$    & $2.4 \times 10^3$    & $1.8 \times 10^3$    & $2.4 \times 10^3$    \\
        \bottomrule
      \end{tabular}
    }
  \end{center}
  \label{table:metallicity_cemp_times}
\end{table}

The time for each species to reach equilibrium increases with
decreasing metallicity. This is a direct consequence of the mass-action kinetics
we used to determine reaction rates. The carbon-enhanced models show faster
timescales for the same reason.

Another interesting investigation involves the pathways in which molecular species
are formed (and disassociated), and how these change throughout the
atmosphere. To pursue this, we represent the reaction network as a weighted
directed graph, as shown in Sec.~\ref{subsec:gcrn}. The nodes are the left and
right sides of the reactions (hereafter complexes), and the edges represent the reactions themselves, weighted by their
corresponding inverse rate. As this graph is often disconnected, and because we are interested in the
species, we added nodes for each individual chemical species.
To connect the graph fully, the individual species nodes have an unweighted edge
to each complex that contains it. In this way, we can represent the evolution of
one species into another by means of the reaction pathways.

We used pathfinding algorithms to move from a source species to a
target species, identifying the chemical pathway and its corresponding timescale
(simply the sum of the edge weights). These change not only with the
temperature, but also with the number densities of the reactants, meaning that the
most frequented pathways for a given source-target pair can change during the
chemical evolution.

The custom pathfinding algorithm (based on Dijkstra's shortest-path algorithm
\citep{dijkstraNoteTwo1959} and taking inspiration from A* pathfinding
\citep{foeadSystematicLiterature2021}) is described in the following steps:

\begin{enumerate}
  \item{Start on a species source node.}
  \item{If the current node is the target node, return the path.}
  \item{Otherwise, find all nodes connected to the current node.}
  \item{If the last travelled edge had a weight of zero, omit all edges with
        weights of zero from the next choice of edges.}
  \item{Pick an edge at random and repeat from 2.}
  \item{Pathfinding will converge once all possible paths from source to target
        have been explored.}
\end{enumerate}

Step $4$ is necessary to prevent species-species jumps that are included as
a side effect of adding chemical species to the graph. These unweighted edges
are represented with a weight of $0$, and traversing two of these consecutively
is unphysical (e.g. moving from CO -> CO + H -> H) as it represents a species
transforming into another (often completely unrelated) species without a
reaction. However, these connections are still necessary to fully
connect the graph; we removed the ability to travel along these connections
consecutively, effectively altering the graph during pathfinding.

In our network, we investigated key pathways from C and O to CO, as well as the
reverse. In all cases, reducing the metallicity resulted in
longer timescales for reactions. Additionally, a single reaction dominates in most pathways and is often referred to as the ``rate-limiting
step'' \citep{tsaiVULCANOpensource2017, tsaiConsistentModeling2018}.
Table~\ref{table:rate_limiting} shows the main reactions involved in the
formation and dissociation of CO for the AM3 atmosphere. We qualitatively
reproduce the same effects as those explored in
\citet{wedemeyer-bohmCarbonMonoxide2005}, and find that of the three reactions
that dissociate CO to C and O, the reaction CO $\rightarrow$ CO + H is by far
the most efficient, even in this extremely metal-poor atmosphere. Additionally,
formation via species exchange (especially by OH) is the most preferable set of
pathways.

\begin{table}[h!]
  \caption{Step-by-step reactions and rate-limiting steps for the AM3 model
    atmosphere at a temperature of $3500$ [K] and a gas density of $10^{-9}$
    [g cm$^{-3}$]. The rate-limiting step (longest step in a pathway) is
    highlighted in bold.}
  \begin{center}
    \scalebox{0.90}{
      \begin{tabular}{l r l c r l }
        \toprule
        Pathway            & Step & Reactants &               & Products  & Timescale [s]                  \\
        \midrule
        C $\rightarrow$ CO &      &           &               &           &                                \\
        \midrule
        \textbf{Pathway 1} & 1.   & C + OH    & $\rightarrow$ & CO + H    & $\mathbf{7.43 \times 10^{-5}}$ \\
                           &      &           &               & Total:    & $7.43 \times 10^{-5}$          \\
                           &      &           &               &           & ---------------                \\
        \textbf{Pathway 2} & 1.   & C + OH    & $\rightarrow$ & CH + O    & $\mathbf{4.10 \times 10^{-3}}$ \\
                           & 2.   & CH + O    & $\rightarrow$ & CO + H    & $4.27 \times 10^{-4}$          \\
                           &      &           &               & Total:    & $4.53 \times 10^{-3}$          \\
                           &      &           &               &           & ---------------                \\
        \textbf{Pathway 3} & 1.   & C + NO    & $\rightarrow$ & CN + O    & $\mathbf{1.44 \times 10^{1}}$  \\
                           & 2.   & CN + O    & $\rightarrow$ & CO + N    & $6.46 \times 10^{-3}$          \\
                           &      &           &               & Total:    & $1.44 \times 10^{1}$           \\
        \midrule
        O $\rightarrow$ CO &      &           &               &           &                                \\
        \midrule
        \textbf{Pathway 1} & 1.   & CH + O    & $\rightarrow$ & CO + H    & $\mathbf{4.27 \times 10^{-4}}$ \\
                           &      &           &               & Total:    & $4.27 \times 10^{-4}$          \\
                           &      &           &               &           & ---------------                \\
        \textbf{Pathway 2} & 1.   & CH + O    & $\rightarrow$ & C + OH    & $\mathbf{2.63 \times 10^{-3}}$ \\
                           & 2.   & C + OH    & $\rightarrow$ & CO + H    & $7.43 \times 10^{-5}$          \\
                           &      &           &               & Total:    & $2.70 \times 10^{-3}$          \\
                           &      &           &               &           & ---------------                \\
        \textbf{Pathway 3} & 1.   & O + C2    & $\rightarrow$ & C + CO    & $\mathbf{6.72 \times 10^{0}}$  \\
                           &      &           &               & Total:    & $6.72 \times 10^{0}$           \\
        \midrule
        CO $\rightarrow$ C &      &           &               &           &                                \\
        \midrule
        \textbf{Pathway 1} & 1.   & CO + H    & $\rightarrow$ & C + OH    & $\mathbf{6.26 \times 10^{-5}}$ \\
                           &      &           &               & Total:    & $6.26 \times 10^{-5}$          \\
                           &      &           &               &           & ---------------                \\
        \textbf{Pathway 2} & 1.   & CO + H    & $\rightarrow$ & C + O + H & $\mathbf{5.27 \times 10^{-1}}$ \\
                           &      &           &               & Total:    & $5.27 \times 10^{-1}$          \\
                           &      &           &               &           & ---------------                \\
        \textbf{Pathway 3} & 1.   & CO + M    & $\rightarrow$ & C + O + M & $\mathbf{5.24 \times 10^{0}}$  \\
                           &      &           &               & Total:    & $5.24 \times 10^{0}$           \\
        \midrule
        CO $\rightarrow$ O &      &           &               &           &                                \\
        \midrule
        \textbf{Pathway 1} & 1.   & CO + H    & $\rightarrow$ & C + OH    & $6.26 \times 10^{-5}$          \\
                           & 2.   & C + OH    & $\rightarrow$ & CH + H    & $\mathbf{4.10 \times 10^{-3}}$ \\
                           &      &           &               & Total:    & $4.16 \times 10^{-3}$          \\
                           &      &           &               &           & ---------------                \\
        \textbf{Pathway 2} & 1.   & CO + H    & $\rightarrow$ & C + O + H & $\mathbf{5.27 \times 10^{-1}}$ \\
                           &      &           &               & Total:    & $5.27 \times 10^{-1}$          \\
                           &      &           &               &           & ---------------                \\
        \textbf{Pathway 3} &      &           &               &           &                                \\
                           & 1.   & CO + C    & $\rightarrow$ & C2 + O    & $\mathbf{3.24 \times 10^{1}}$  \\
                           &      &           &               & Total:    & $3.24 \times 10^{1}$           \\
        \bottomrule
      \end{tabular}
    }
  \end{center}
  \label{table:rate_limiting}
\end{table}

We examined the preferred
pathways in the network for OH for three abundance mixtures: AM3, AC1, and AC2. AM3
and AC2 are qualitatively similar, where radiative association of OH via H is
a leading timescale. Species exchange with CH and CO is not as preferable. AC2
shows exactly the opposite trend, with species exchange routes being
significantly better travelled than direct radiative association. Again, this is
because in both AM3 and AC2, more free O is available after CO has been formed,
while in AC1, very little O is present and OH formation relies on carbonic
species.

\subsection{Treatment of photochemistry}
\label{subsec:photochemistry}

Our network does not include the effects of photodissociation of species because of the greatly increased complexity required to treat this process properly. In
the collision-dominated layers, photochemistry is unlikely to be important, but
the situation may be different in higher optically thin layers, where
radiation-driven processes become important. The importance of photochemistry is
perhaps traced better by the prominence of radiative NLTE effects. The treatment
of neutral C in the Sun \citep{amarsi3DNonLTE2019} and O in the Sun
\citep{steffenPhotosphericSolar2015} shows that the abundances are affected up
to $0.1$ dex in relevant line-forming regions. It is feasible that
photochemistry is then an important consideration in higher layers, but
the treatment of the photochemical reactions of all atomic and molecular species is a
considerably difficult and time-consuming endeavour. We welcome any further
advancements in this direction.
\subsection{Complexity reduction}
\label{subsec:complexity_reduction}

Ideally, we would like to include as many species and reactions as possible
into the network to model it as precisely as possible. Unfortunately, due to the
large memory cost of storing 3D arrays as well as the steep scaling of the
solution time with the size of the kinetic system, methods that reduce complexity
are often required. In this work, we have presented a heavily reduced network
that is focused on the formation and disassociation of a few key molecular
species. However, the existence and addition of other species into the network
can alter evolution, pathways, and timescales. It is often the case that only a
small subset of reactions controls the vast majority of the evolution.
Identifying these reactions can prove challenging, but a few methods exist to
reduce the complexity of the kinetics problem
\citep{grassiComplexityReduction2012,popeComputationallyEfficient1997}.
In our case, the network was already heavily reduced to the key reactions, and
chemical pathways were investigated by \citet{wedemeyer-bohmCarbonMonoxide2005},
which in part verify this. In the future, we aim to investigate chemical
pathways found via a graph theoretical analysis to reduce the number of
reactions and species to only those necessary to model significant trends in
the regions of interest.
\subsection{Potential error from LTE assumptions}
\label{subsec:lte_error}

In the analysis presented thus far, several assumptions have been made that can
introduce small errors into the process. It has been demonstrated that the
assumption of chemical equilibrium generally holds well in the photospheres of
the stars considered in this work. However, deviations increase in higher
layers, and several assumptions made about these higher layers should be
addressed. Since chemical feedback is
neglected, chemical reactions do not affect the temperature of the model
atmosphere directly, despite being exo- or endothermic. Furthermore, the
equation of state does not take into account the departures from chemical
equilibrium. Finally, radiative NLTE corrections may be significant in the
higher layers of the atmospheres we consider.

Firstly, the absence of heating and cooling by chemical reactions contributes
only a small error. Here we consider the formation of C and O into CO because CO has
the highest bond energy of any species presented here and will naturally
dominate the effects of chemical feedback due to this property and its
abundance. As an extreme case, we considered the conversion of all C into CO
in the AM1 atmosphere. This atmosphere is the most metal rich and is
oxygen rich, so that this conversion provides an upper limit to the expected change
in heat and temperature. The specific heat per unit mass is nearly constant in
the upper layers of this atmosphere, with
$c_\mathrm{V}\approx 10^8$~erg~g$^{-1}$~K$^{-1}$. The formation of CO releases
$e_\mathrm{diss}:=11.16$~eV ($1.79\times 10^{-11}$~erg) of energy per reaction,
and the energy released per mol is found by multiplying by Avogadro's number. The
number fraction of C (where the total amount of species is normalised to unity)
divided by the mean molecular weight $\mu := 1.26$ in the neutral stellar atmosphere
gives $A_\mathrm{C} \approx 2.4\times10^{-4}$. We therefore find the change in
temperature
\begin{equation}
  \Delta T = \frac{e_\mathrm{diss}~N_\mathrm{A}~A_\mathrm{C}}{\mu~c_\mathrm{V}} \approx 20~\mathrm{K},
\end{equation}
where $N_\mathrm{A}$ is Avogadro's number.
In the lower-metallicity models presented here, the effect would be
significantly smaller. The lack of chemical
feedback therefore adds a negligible error to the total energy and
temperature.

Secondly, the effects of deviations from equilibrium number densities could have
a small effect on the overall structure, but the deviations shown here are
relatively small and are generally confined to trace species. The mean molecular
weight is dominated by H and He (metals are around $\sim 1\%$), the
deviations of which are completely negligible. Any deviations
on the scale seen here would therefore not have
significant effects on the overall structure of the model atmosphere.

Thirdly, NLTE radiative transfer may be important when considering chemical
kinetics in highly optically thin layers.
\citet{popaNLTEAnalysis2022} showed the importance of this feature on the CH
molecule in metal-poor red giant atmospheres, which suggests an increase in
measured C abundance compared to LTE techniques (increasing with decreasing
metallicity). We find a small increase in CH number density compared to
equilibrium values, which would offset the large increase in the LTE carbon
abundance required to reproduce the NLTE spectra. Additionally, when
the collisional dissociation rates of CH are compared (e.g. Reaction $1$), the rates are
comparable to or higher than the photodissociation rates across the optically
thin atmosphere. We estimated the photodissociation rate of CH with the help of
the continuous opacity of provided by \citet{kuruczOHCH1987}. For the population
numbers, LTE was assumed, and the radiation field of the 1D atmosphere was
calculated with opacity distribution functions. The obtained rate suggests
overall that photodissociation does not completely dominate the process.
It is therefore interesting to
consider the interplay between these processes, and we look forward to
additional efforts in this field.

Finally, it should be stated that the models presented in this work do not
include a comprehensive treatment of the stellar chromosphere.
While we note that deviations from chemical equilibrium are
important in stellar chromospheres, further work in these areas is therefore necessary and
welcome.
\section{Conclusion}
\label{sec:conclusion}

We have presented a study of 3D time-dependent molecular formation and
dissociation in one solar metallicity and four metal-poor atmospheres. The
chemistry was modelled through
mass-action kinetics with $\nrxns$ reactions and $\nspecies$ species that are
advected by the velocity field during the hydrodynamics step. We additionally
presented a comparison to the equilibrium abundances, computed with Python or Julia
chemical kinetics codes. Deviations from equilibrium are seen primarily in higher
photospheric
layers, around shocks, and in the temperature differences throughout convection
cells.

\begin{itemize}
  \item{
        In all models presented in this work, molecular species are
        generally in chemical equilibrium throughout the model photospheres.
        Molecular species show mean deviations from equilibrium reaching
        $0.15$ in the lower chromosphere, and these deviations increase with
        decreasing metallicity and increasing height. The largest deviations
        are in CN, \CT\ , and CH when $\log$ (C/O) $<1$, and in OH when
        $\log$ (C/O) $>1$. Above $\log{\tau} \approx -2$, the less abundant molecule
        of C or O becomes locked into CO, inhibiting the formation of other
        molecular species involving that species. This results in
        comparatively low amounts of CH, CN, and \CT\ in all models except
        AC2, and comparatively low amounts of OH in model AC2.
        }
  \item{
        The deviations from equilibrium can also be attributed to
        the behaviour around chromospheric shock waves. In the equilibrium case,
        the hot shock front contains very low number densities of molecular
        species, while the time-dependent treatment has greater number
        densities as the evolution proceeds with a finite timescale. In the
        uppermost coolest layers (T $\leq \sim 3500$ [K]), slow chemical
        timescales result in a depletion of CO as there is insufficient time
        to form it before material is advected to a significantly different
        thermodynamic state.
        }
  \item{These deviations are unlikely to contribute significantly to
        spectroscopic measurements for metal-poor dwarfs because the line cores
        of key molecular species are generally
        formed in deeper layers \citep{gallagherIndepthSpectroscopic2017}.
        The largest deviations are mostly outside of the range of
        the contribution functions for the CH G-band and OH-band,
        but these deviations could still affect spectral line shapes, which
        can only be properly reproduced in 3D models. The perceived trend of
        increased carbon enhancement with decreasing stellar metallicity is
        therefore not due to an improper treatment of time-dependent
        chemistry. An investigation
        including spectrum synthesis using the time-dependent number
        densities is however warranted in light of these deviations.
        }
  \item{
        Relative deviations increase with
        decreasing metallicity due to slower mass-action reaction rates.
        The change in metallicity does not lead to a strictly linear
        increase in chemical timescale or decrease in yield in all layers,
        but generally, lower metallicities result in longer chemical
        timescales and lower yields.
        }
  \item{
        The C/O ratio plays a key role in determining which
        molecular species are further out of equilibrium. Both CH and OH are
        formed along reaction
        pathways to form CO. In the majority of atmospheres we presented,
        oxygen is present in excess compared to carbon, making OH formation
        more viable than CH. This leads to faster chemical timescales for
        reaction pathways involving OH. Changing this ratio so that carbon is in
        excess likewise changes the pathways to make the formation of
        carbon-bearing species preferential.
        }
  \item{
        The lack of chemical feedback contributes a negligible        error to the evolution of chemical species, energy, and momentum of
        the system because these metals are trace species. NLTE radiative
        transfer has been shown to be of importance in higher layers
        \citep{popaNLTEAnalysis2022}, and our deviations point in the
        opposite direction to those seen through radiative NLTE
        calculations. Some kinetic rates are as fast or faster than our
        calculated photodissociation rates across the optically thin
        atmosphere, suggesting that both radiative NLTE transfer and
        time-dependent kinetics are important to consider in these layers.
        }
\end{itemize}

In conclusion, we find molecular species to generally be in a state of chemical
equilibrium under photospheric conditions for the models presented in this work.
The effect of altering the C/O ratio is directly seen in the final yields of
molecular species such as CH and OH. While relative deviations increase with
decreasing metallicity because the mass-action kinetic rates are slower, these effects
are not large enough to contribute significantly to spectroscopic abundance
measurements because the line cores of interest for these stars are generally formed
in deeper regions in which chemical equilibrium is well established. The deviations
increase with height, and it is likely that there is interesting interplay
between radiative and kinetic departures from LTE in the upper photospheres and
chromospheres of metal-poor stars.

\begin{acknowledgements}
  S.A.D. and H.G.L. acknowledge financial support by the Deutsche
  Forschungsgemeinschaft (DFG, German Research Foundation) -- Project-ID
  138713538 -- SFB 881 (``The Milky Way System'', subproject A04).
\end{acknowledgements}

\bibpunct{(}{)}{;}{a}{}{,} 
\bibliographystyle{aa}
\bibliography{citations.bib}

\begin{appendix}
  %
  \section{Carbon enhancement and metallicity}
  \label{app:ce_metallicity}

\begin{figure}[h]
    \centering
    \scalebox{0.8}{
      \includegraphics[]{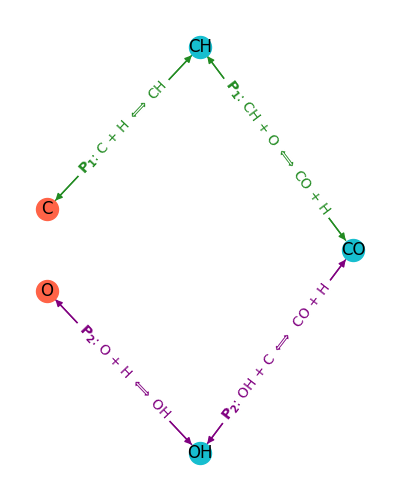}
    }
    \caption{Diagram of the simplified reaction network showing pathways
      $P\subrm{1}$ and $P\subrm{2}$.}
    \label{fig:pathfinding_network}
  \end{figure}

  The effect of the C/O ratio can be visualised by investigating
  reaction pathways between the species C, O, CH, OH, and CO.
  A simplified reaction network with eight reactions was used in this
  study, and the temperature dependence was removed from the rate coefficients to
  investigate the effects of mass-action kinetics. This simplified network takes a
  subset of reactions from the original network used in this work that
  characterise the formation and dissociation of CO via CH and OH.
  In this network, there are $\text{two}$ pathways from C and O to CO (and back), shown
  and labelled in Fig. \ref{fig:pathfinding_network}.
  The pathway involving C-CH-CO is labelled $P\subrm{1}$ and that of O-OH-CO is labelled
  $P\subrm{2}$. Additionally, the pathways
  are constructed such that pathways $P\subrm{1}$ and $P\subrm{2}$ have the same
  rates when the amount of C and O is equal. This symmetry allows us to
  investigate the effect of the C/O ratio alone on the mass-action
  kinetics. The reaction rates are constant in temperature, but are otherwise
  constructed to qualitatively match those of the larger network (apart from the
  symmetry).

\begin{figure}[h]
    \centering
    \scalebox{0.7}{
      \includegraphics[]{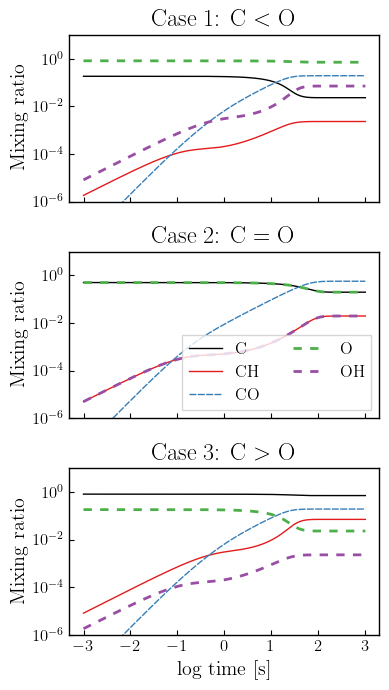}
    }
    \caption{Evolution of C, CH, CO, O, and OH for a simplified reaction network
      we used to illustrate the effect of the C/O ratio on equilibrium timescales.}
    \label{fig:pathfinding_experiment}
  \end{figure}

  Fig. \ref{fig:pathfinding_experiment} shows the evolution of these species for
  three cases. Case 1 contains more oxygen than carbon. In case 2, the amount
  of carbon and oxygen is equal. Case 3 contains more carbon than oxygen. As
  expected, the overall yield of the number densities reflects that of the input
  abundances.

  We adopted abundances similar to the AM2 model for the $\text{three}$ cases.
  The total timescale was calculated by summing the individual timescales
  of the reactions along the pathway. The total timescales and abundances
  considered for each case are shown in Table
  \ref{table:pathfinding_timescales}. The shortest timescales are highlighted in
  bold. These are not the same timescales as shown in
  Table~\ref{table:metallicity_cemp_times}; the timescales shown there represent
  the times various species evolve to chemical equilibrium, while the timescales
  here are inverse reaction rates scaled by number densities. In this sense, they
  are essentially e-folding timescales.

  \begin{table}[h!]
    \caption{Total timescales for the C-CH-CO pathway $P\subrm{1}$ and the
      O-OH-CO pathway $P\subrm{2}$ (towards formation of CO) for three different
      $\log$ C/O ratios.}
    \begin{center}
      \begin{tabular}{l c c r r r }
        \toprule
        Case & A(C)   & A(O)   & $\log$ C/O & $P\subrm{1}$   & $P\subrm{2}$   \\
             &        &        &            & [s]            & [s]            \\
        \midrule
        1    & $6.41$ & $7.06$ & $-0.65$    & $36.8$         & $\mathbf{1.6}$ \\
        2    & $7.06$ & $7.06$ & $0.00$     & $3.6 $         & $3.6$          \\
        3    & $7.06$ & $6.41$ & $+0.65$    & $\mathbf{1.6}$ & $36.8$         \\
        \bottomrule
      \end{tabular}
    \end{center}
    \label{table:pathfinding_timescales}
  \end{table}

  In case 1, where O is more abundant than C, pathway $P\subrm{2}$ is faster because
  forming OH is favourable to forming CH. This results in C and CH being
  considerably depleted compared to the other species, and it is the same effect as
  in model AM2. In case 2, where O and C are equally abundant, CH and OH form
  equally
  quickly due to the symmetry in the simplified network. In case 3, where C is
  more abundant than O, the exact opposite behaviour to case 1 is observed (again
  due to the symmetry), but this is qualitatively similar to model AC2. All in
  all, the current method allows us to investigate favourable pathways in a given
  network, validating the chemical species that are especially important when
  considering the evolution of another. The longer pathways seen here in
  the simplified network are indicative of similar trends in the full network,
  reinforcing the notion that as chemical timescales grow longer, approaching
  dynamical timescales, it becomes increasingly likely that chemical species lie
  farther from their equilibrium values. However, due to the inclusion of more
  species and many more pathways between them, the dynamics of the full network
  are significantly more complicated, and it is not immediately clear from the
  timescales alone how far species will be out of equilibrium, nor whether they
  would be present in excess or in depletion. Nevertheless, an analysis of
  favourable reaction pathways might lead to improvements such as complexity
  reduction.
  \section{Slices through the photosphere}
  \label{app:extra_figs}

  The following figures show horizontal and vertical slices through the
  photospheres of all models, analogously to Figs.~\ref{fig:mm30_xy} and
  \ref{fig:mm30_xz}. The slices in the $xy$ direction are all taken at an optical
  depth of $\log \tau = -4$. In all cases, the differences in the $xy$ slices are
  minor, seen primarily around hot shock fronts where equilibrium chemistry
  predicts nearly instantaneous dissociation of molecular species, while in the
  time-dependent case, this proceeds on a finite timescale. The  time-dependent molecular species are therefore seen in excess.

  In the $xz$ slices, differences are seen in the uppermost layers of the
  atmosphere (generally only in chromospheric layers). These layers are frequented
  by shock waves that disrupt molecular formation, and when considering
  equilibrium chemistry, the effects of molecular dissociation
  again occur too quickly in the hot shock fronts, and formation of CO occurs too
  quickly in the cooler regions. As shown in
  Figs.~\ref{fig:metallicity_comparison} and \ref{fig:cemp_comparison}, molecular
  formation is very slow (on the order of hours to days), and most molecular
  species first show an excess in their yield profile before this decreases as
  they form CO.

  \begin{figure*}
    \centering
    \scalebox{0.67}{
      \includegraphics[]{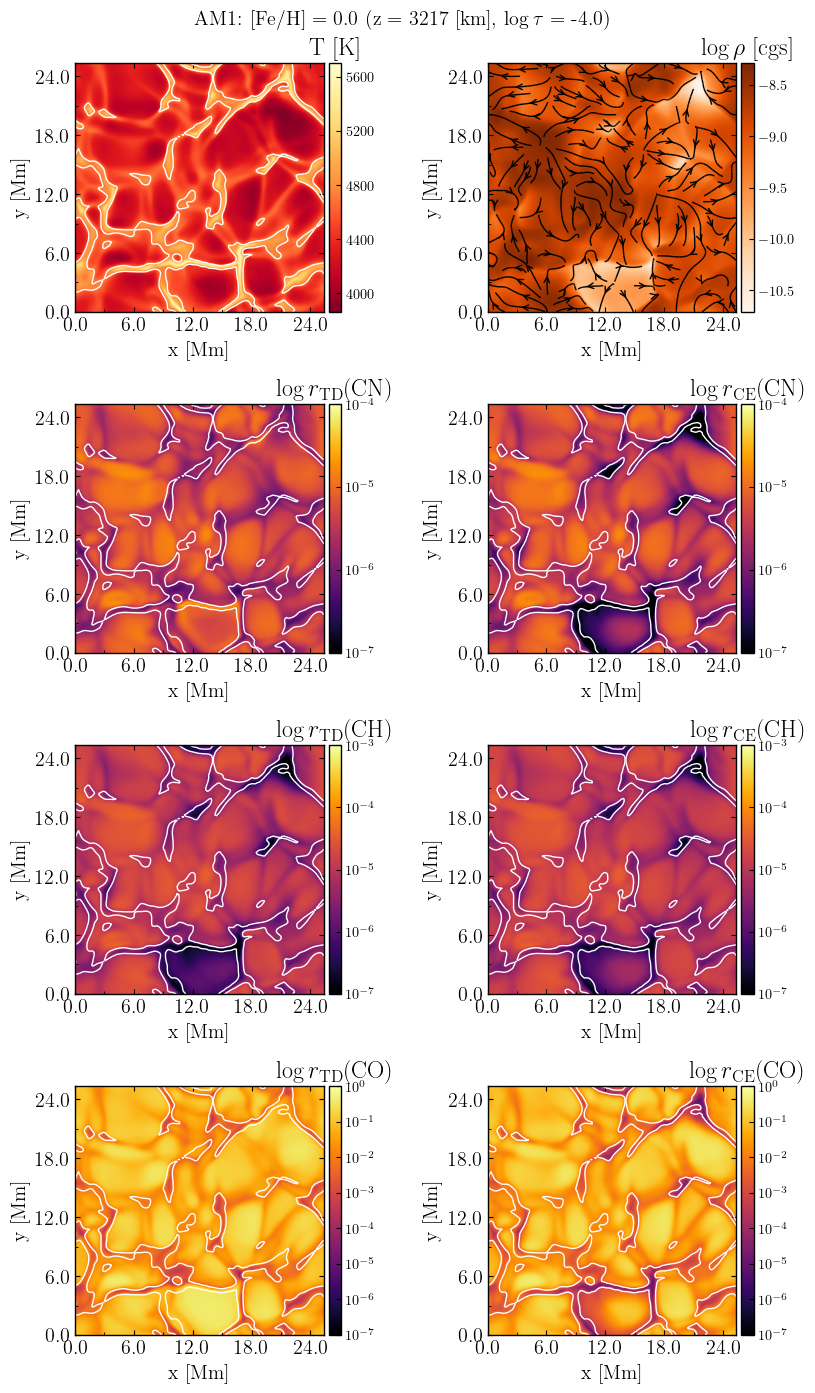}
    }
    \caption{As Fig.~\ref{fig:mm30_xy}, but with [Fe/H] = $+0.0$. The contour
      line traces a temperature of $5000$~K and highlights areas in which deviations
      may be present (seen primarily in CN and CO).}
    \label{fig:mm00_xy}
  \end{figure*}

  \begin{figure*}
    \centering
    \scalebox{0.62}{
      \includegraphics[]{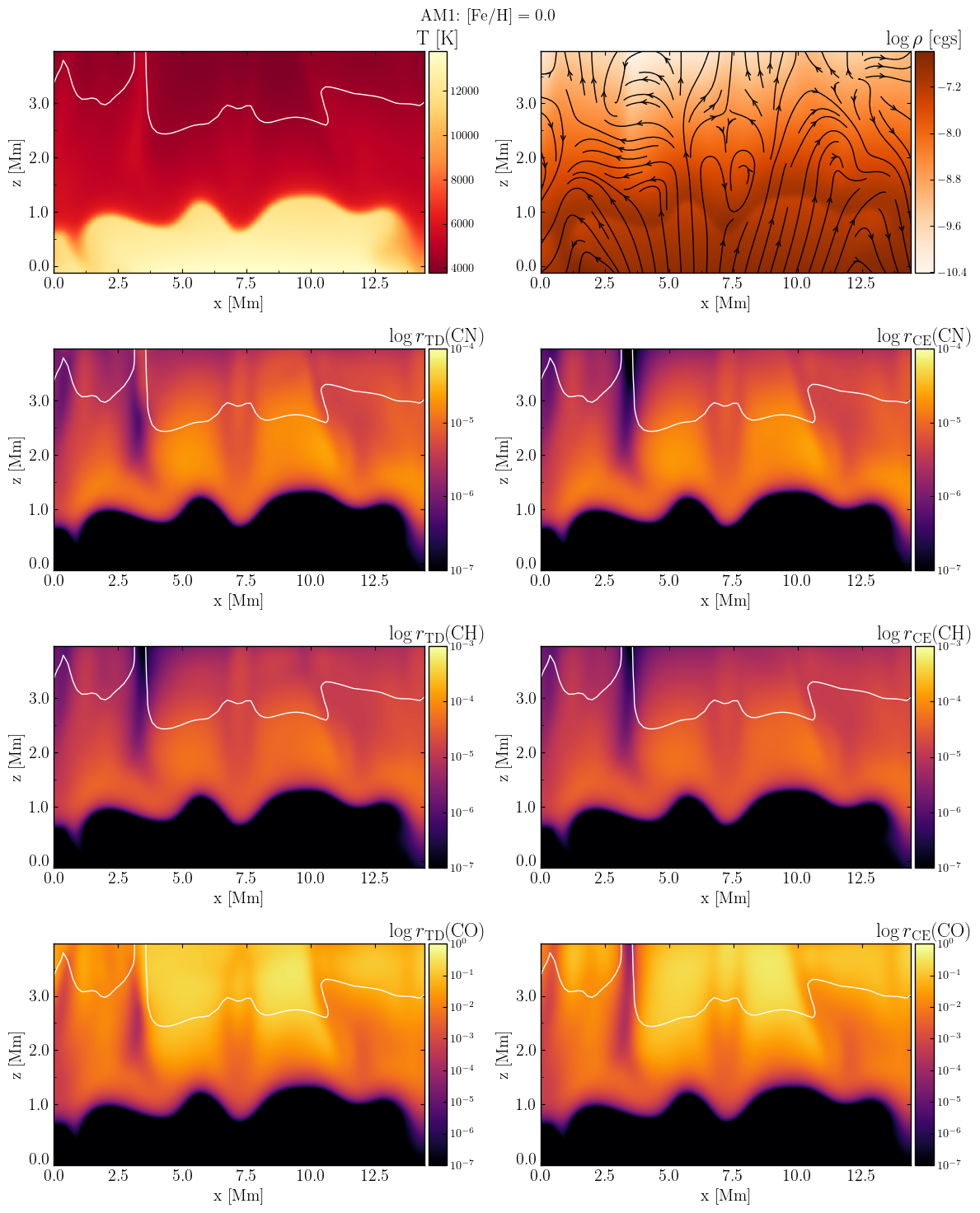}
    }
    \caption{As Fig.~\ref{fig:mm30_xz}, but with [Fe/H] = $+0.0$. The contour
      line traces a temperature of $4550$~K. Minor deviations are present in the
      uppermost layers of the atmosphere for CN and CO.}
    \label{fig:mm00_xz}
  \end{figure*}

  \begin{figure*}
    \centering
    \scalebox{0.67}{
      \includegraphics[]{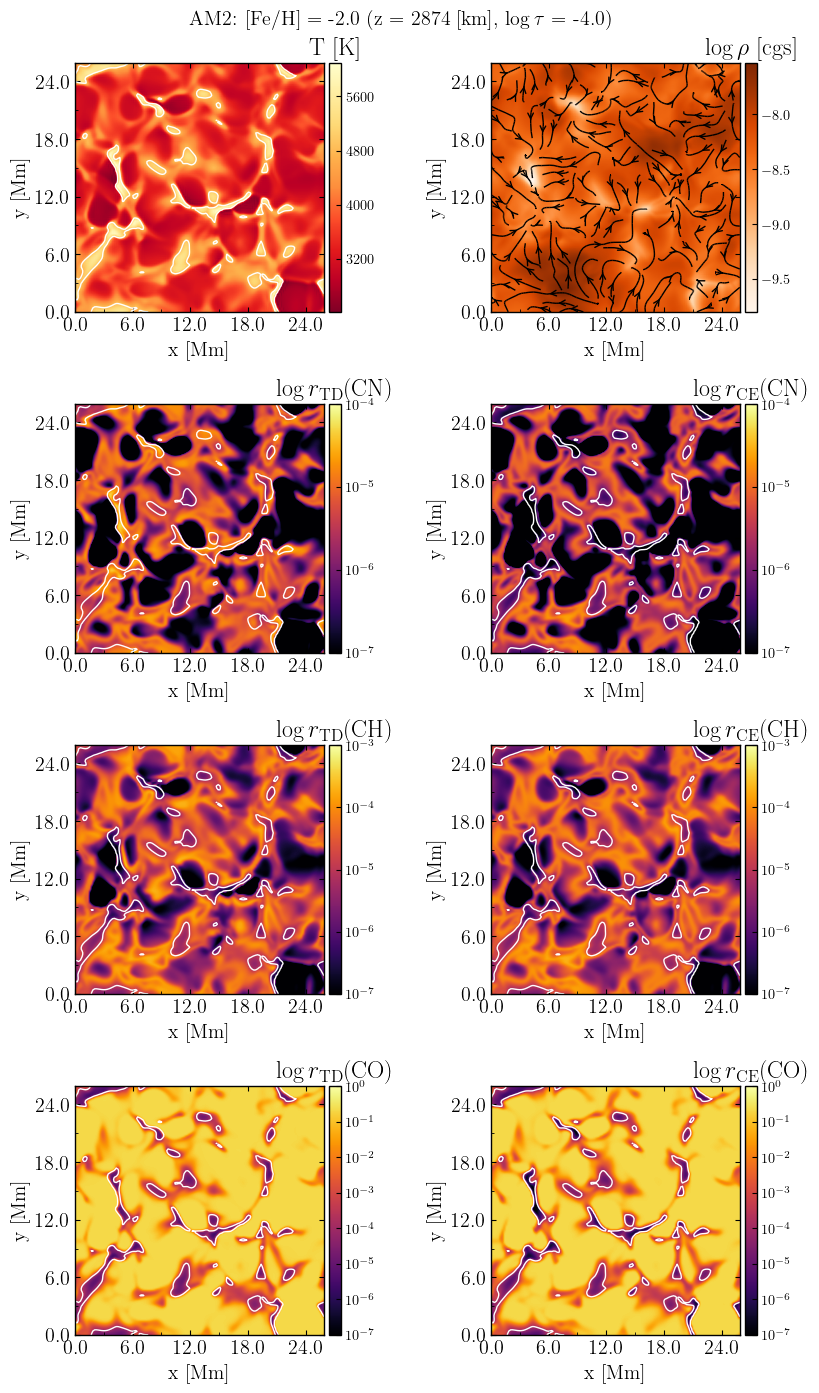}
    }
    \caption{As Fig.~\ref{fig:mm30_xy}, but with [Fe/H] = $-2.0$. The contour
      line traces a temperature of $5000$~K, highlighting areas in which CN chemistry
      is out of equilibrium. CH and CO are generally formed under
      equilibrium conditions.}
    \label{fig:mm20_xy}
  \end{figure*}

  \begin{figure*}
    \centering
    \scalebox{0.62}{
      \includegraphics[]{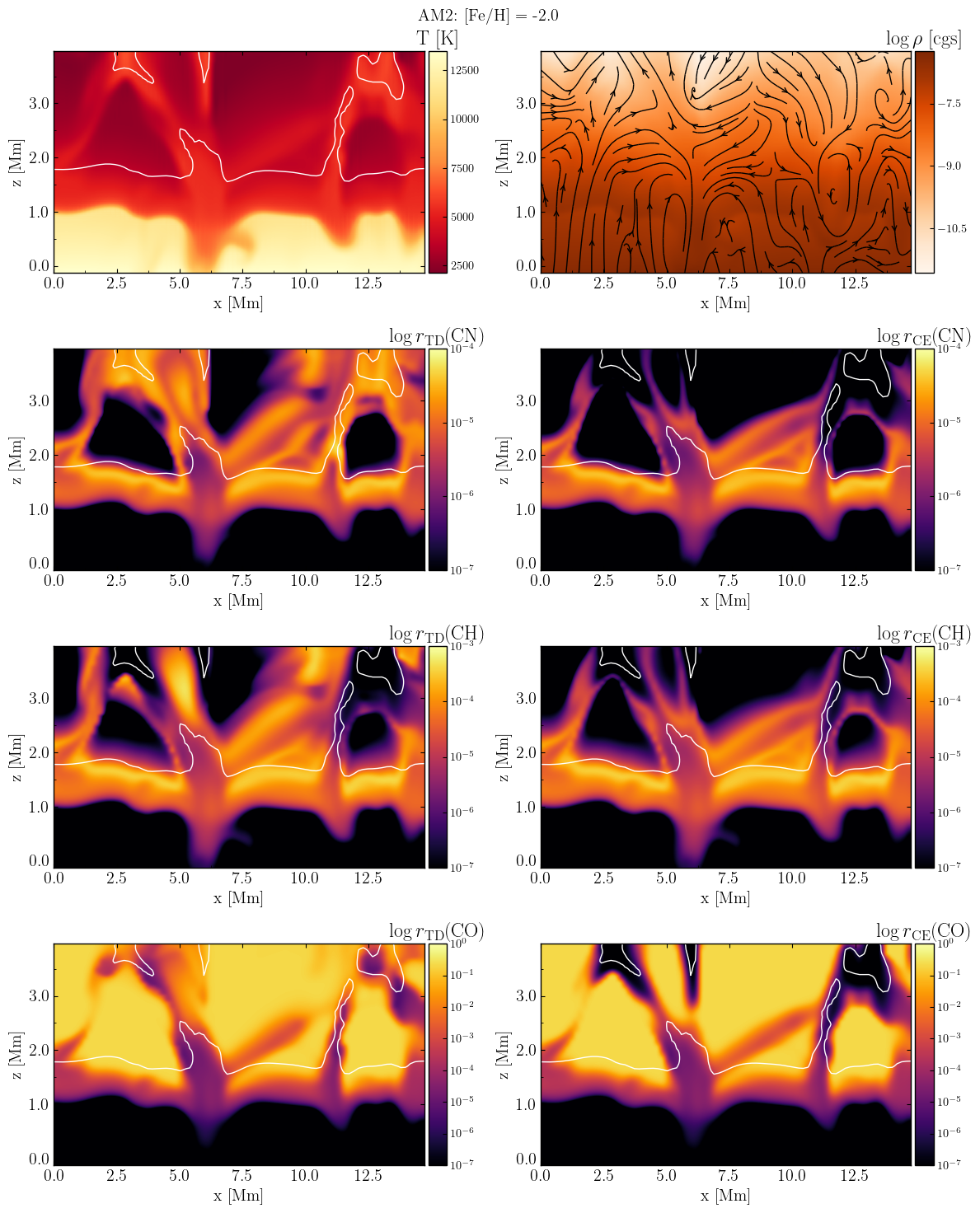}
    }
    \caption{As Fig.~\ref{fig:mm30_xz}, but with [Fe/H] = $-2.0$. The contour
      line traces a temperature of $5400$~K and highlight hydrodynamical features
      such as an updraft colliding with a downdraft around $x = 6.0$~Mm,
      $z = 2.5$~Mm. All molecular species are out of equilibrium around this
      feature as they do not dissociate as quickly as predicted by the equilibrium
      chemistry.}
    \label{fig:mm20_xz}
  \end{figure*}

  \begin{figure*}
    \centering
    \scalebox{0.67}{
      \includegraphics[]{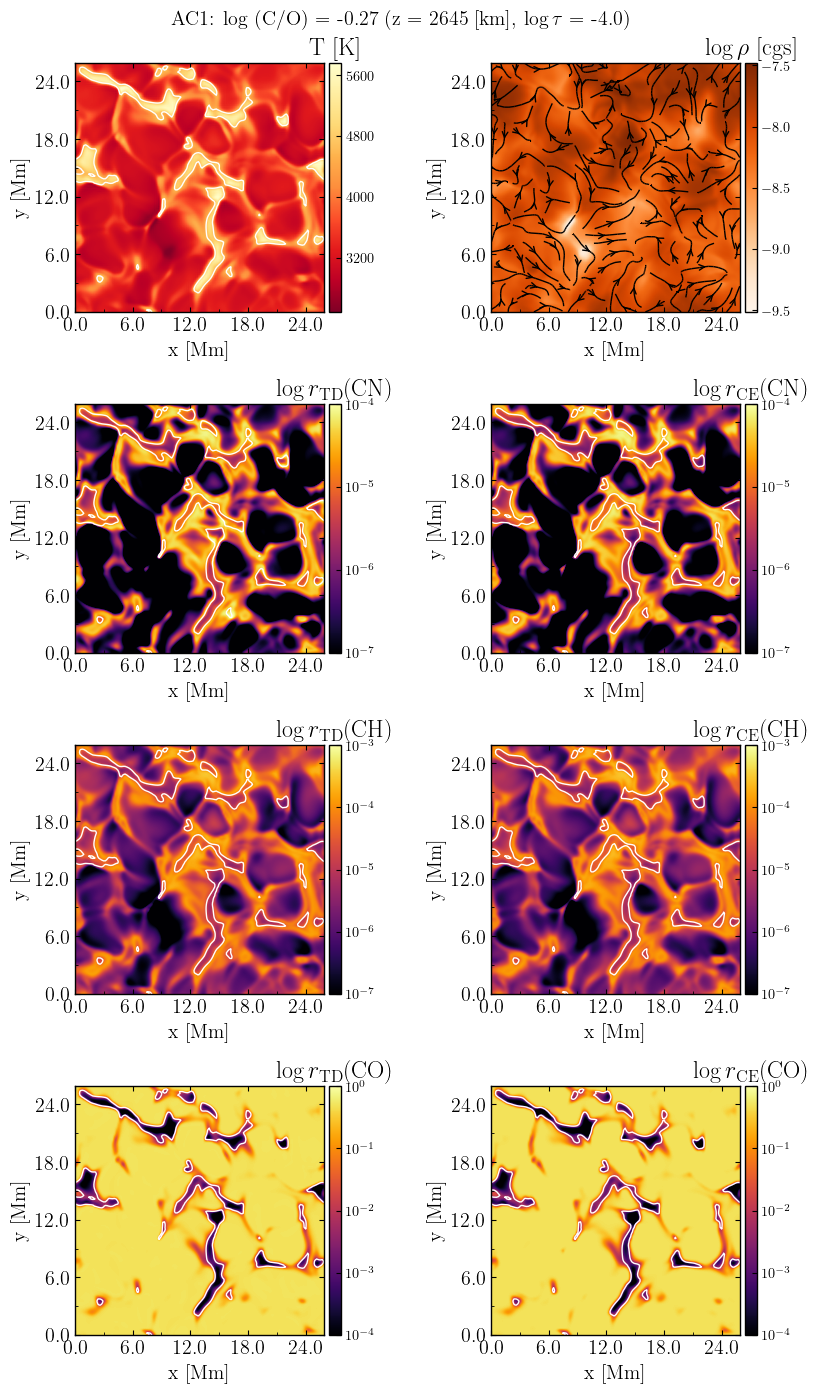}
    }
    \caption{As Fig.~\ref{fig:mm30_xy}, but with a C and O enhancement of
      $+2.0$~dex and C/O = $-0.27$ The contour line traces a temperature of
      $4900$~K. The chemistry is almost entirely in equilibrium in this layer of
      the atmosphere (due to the enhanced C and O abundances), and minor deviations
      are visible only for CN around the hotter regions.}
    \label{fig:ac1_xy}
  \end{figure*}

  \begin{figure*}
    \centering
    \scalebox{0.62}{
      \includegraphics[]{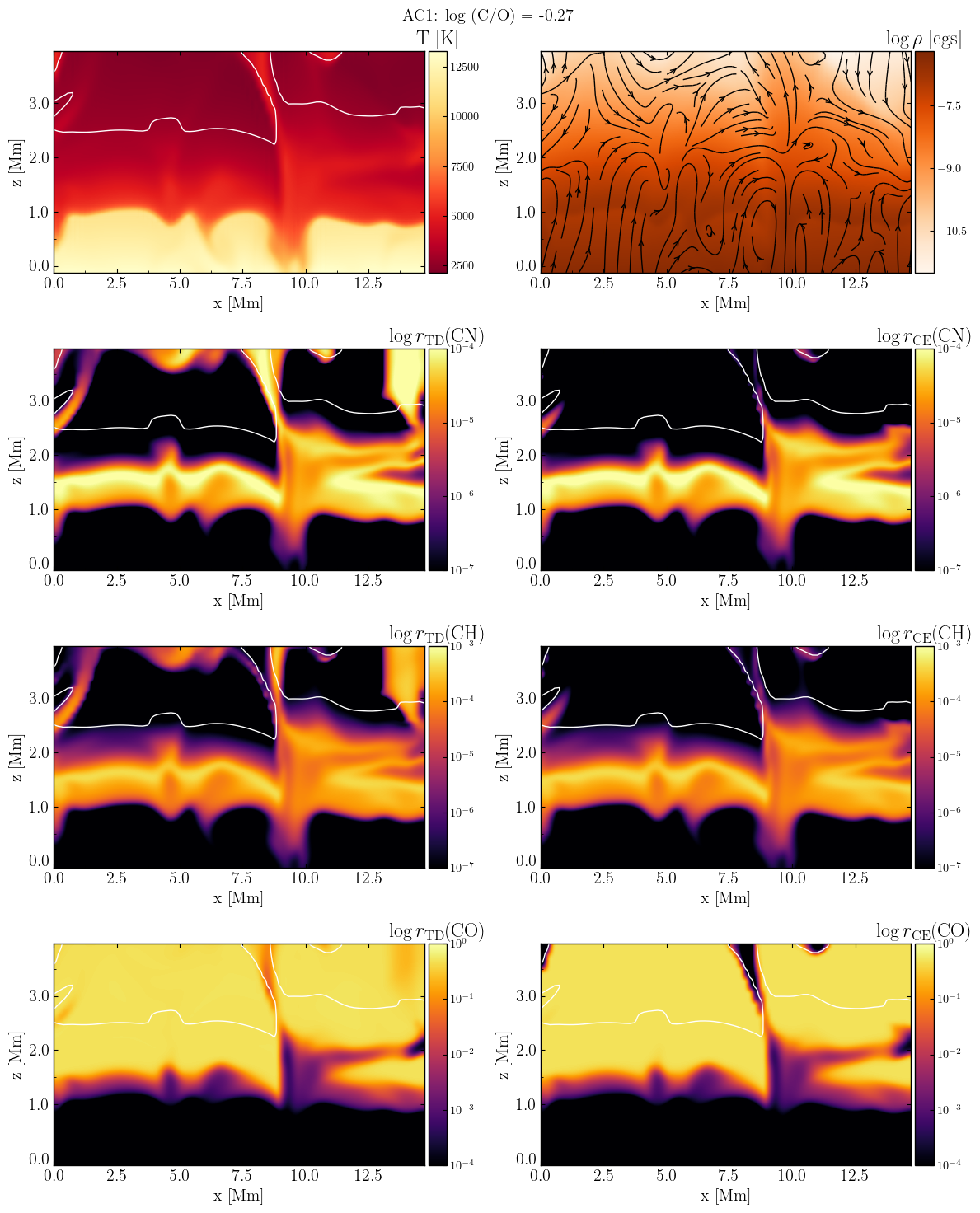}
    }
    \caption{As Fig.~\ref{fig:mm30_xz}, but with a C and O enhancement of
      $+2.0$~dex and C/O = $-0.27$. The contour line traces a temperature of
      $5150$~K and highlights a hot updraft near
      $x = 9.0$~Mm where primarily CN and CO molecular dissociation is out of
      equilibrium.}
    \label{fig:ac1_xz}
  \end{figure*}

  \begin{figure*}
    \centering
    \scalebox{0.67}{
      \includegraphics[]{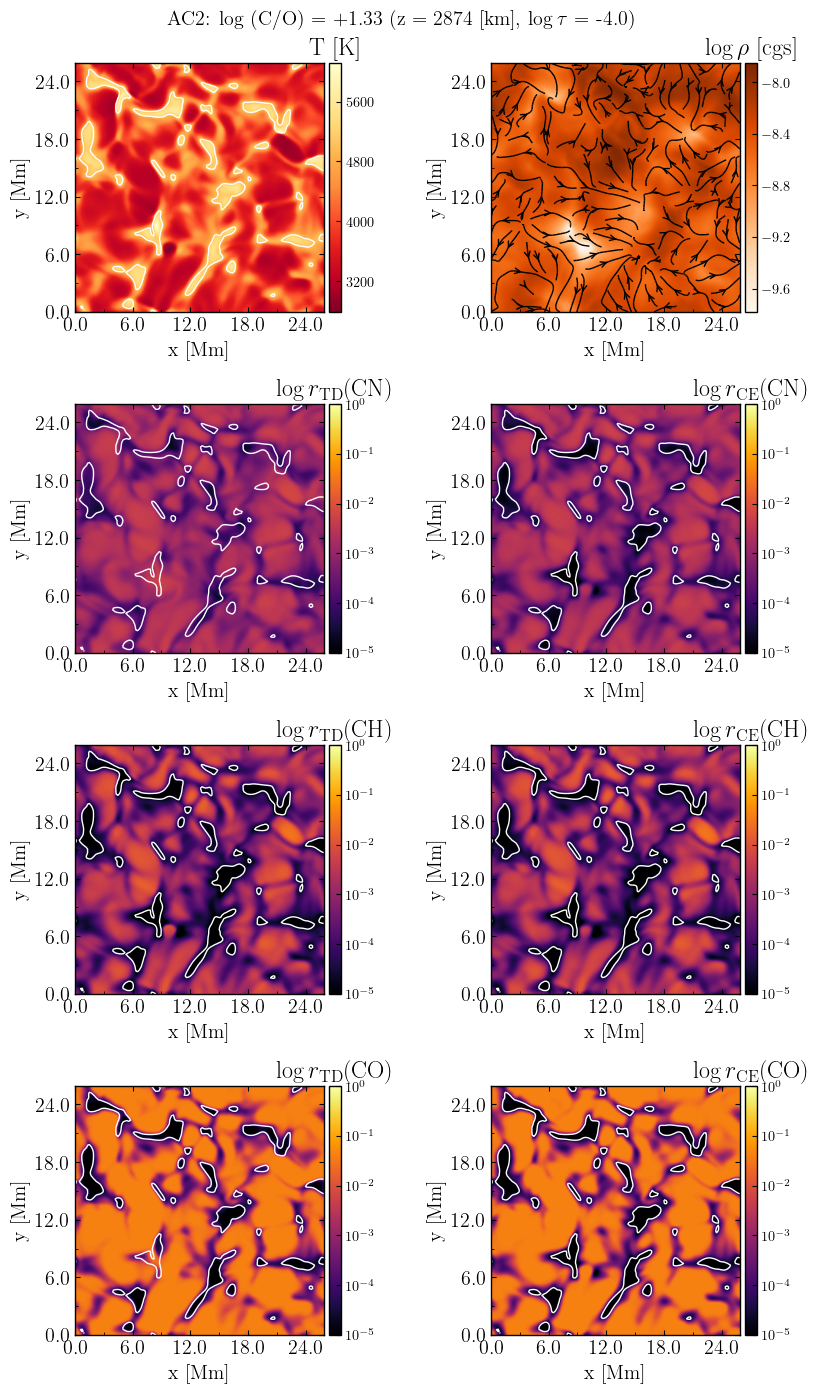}
    }
    \caption{As Fig.~\ref{fig:mm30_xy}, but with a C enhancement of $+2.0$~dex and
      C/O = $+1.33$. The contour line traces a temperature of $5000$~K, showing
      minor deviations present only in CN. The relatively high carbon abundance
      results in the molecular chemistry being very close to equilibrium
      conditions.}
    \label{fig:ac2_xy}
  \end{figure*}

  \begin{figure*}
    \centering
    \scalebox{0.62}{
      \includegraphics[]{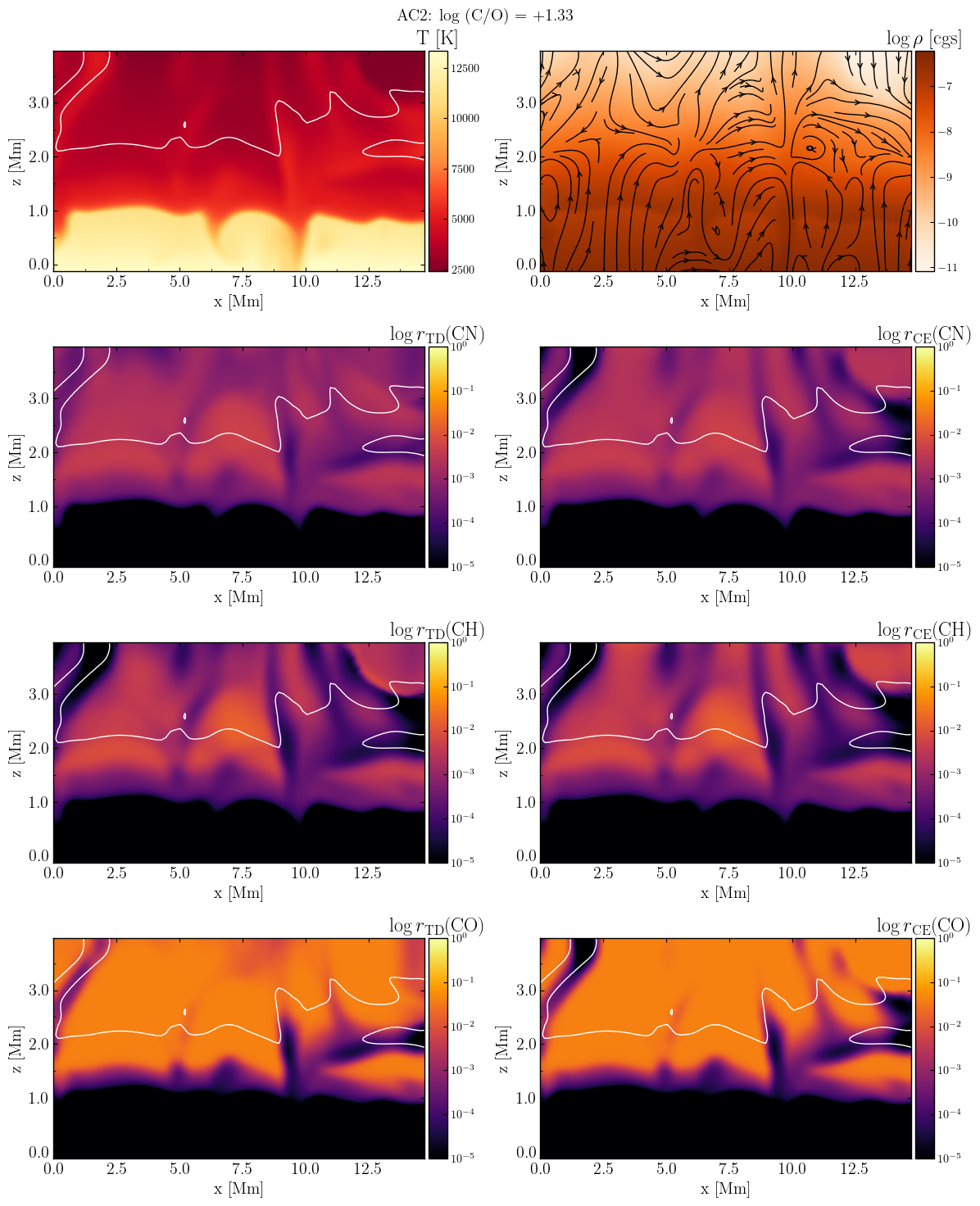}
    }
    \caption{As Fig.~\ref{fig:mm30_xz}, but with a C enhancement of $+2.0$~dex and
      C/O = $+1.33$. The contour line traces a temperature of $4950$~K, and the
      largest deviations are seen in a hot updraft near $x = 6.0$~Mm,
      $z = 3.0$~Mm. The relatively high carbon abundances result in these
      molecular species being formed very close to equilibrium conditions, although
      differences are present in these very diluted regions.}
    \label{fig:ac2_xz}
  \end{figure*}
\end{appendix}
\end{document}